\documentclass{article}
\usepackage{arxiv}
\usepackage[utf8]{inputenc} 
\usepackage[T1]{fontenc}    
\usepackage{hyperref}       
\usepackage{url}            
\usepackage{booktabs}       
\usepackage{cite}
\usepackage{amsmath,amssymb,amsfonts}
\usepackage{algorithmic}
\usepackage{graphicx}
\usepackage{textcomp}
\usepackage{siunitx}
\usepackage{float}
\usepackage{float}
\usepackage{bm}
\makeatletter
\AtBeginDocument{\DeclareMathVersion{bold}
\SetSymbolFont{operators}{bold}{T1}{times}{b}{n}
\SetMathAlphabet{\mathrm}{bold}{T1}{times}{b}{n}
\SetMathAlphabet{\mathit}{bold}{T1}{times}{b}{it}
\SetMathAlphabet{\mathbf}{bold}{T1}{times}{b}{n}
\SetMathAlphabet{\mathtt}{bold}{OT1}{pcr}{b}{n}
\SetSymbolFont{symbols}{bold}{OMS}{cmsy}{b}{n}
\renewcommand\boldmath{\@nomath\boldmath\mathversion{bold}}}
\makeatother

\def\BibTeX{{\rm B\kern-.05em{\sc i\kern-.025em b}\kern-.08em
    T\kern-.1667em\lower.7ex\hbox{E}\kern-.125emX}}
\usepackage[switch]{lineno}

\restylefloat{table}

\title{Comparing time and frequency domain numerical methods with Born-Rytov approximations for far-field electromagnetic scattering from single biological cells}

\author{Cael Warner \\ University of Alberta,  \\ Edmonton, AB T6G2R3 Canada \\
\texttt{spencerw@ualberta.ca}}

\begin{document}
\maketitle

\begin{abstract}
The Born-Rytov approximation estimates effective refractive index of biological cells from measurements of scattered light intensity, polarization and phase. Effective refractive index is useful for estimating a biological cell's dry mass, volume, and internal morphology directly from its elastic light scattering pattern. This work compares the Born-Rytov approximation with analytical, Yee-lattice finite-difference time-domain, and discrete-dipole approximations to Maxwell's equations in the cases of electromagnetic scattering from a sphere and a tomographic reconstruction of Saccharomyces \emph{cerevisiae}. Practical advantages and limitations of each numerical method are compared for modeling electromagnetic scattering of both near-field intensity and the far-field projected intensity, in terms of accuracy, memory, and compute time. When compared with a commercial software implementation of the Yee-lattice finite-difference time domain method, the Born-Rytov scattering approximation and discrete dipole approximation show better agreement with the far-field light scattering pattern from Saccharomyces \emph{cerevisiae}.
\end{abstract}

\keywords{
Single cell \and refractive index \and static light scattering \and computational biology}

\section{Introduction}
\label{sec:introduction}
Characterizing morphological, material, and chemical properties of biological cells is a non-trivial task \cite{leeuwenhoek12observations,herman2020fluorescence}. Properties of biological samples are often inferred from microscopic reconstructions of light intensity, polarization, and/or phase transmitted in the visible spectrum through thin transparent biological samples and spherical lens systems. Kirchoff's theory of diffraction describes light intensity as a scalar solution to the Helmholtz equation \cite{kirchhoff1883theorie}, which Abbe found fails to converge at a focal position \cite{abbe1873ueber}. Spherical abberation imposes an upper-bound cut-off frequency as the diffraction limit for optical images based on light intensity alone. De-focused light intensity, as a consequence of diffraction, varies in phase and polarization over the area of most photon detectors. Light intensity in microscope images is blurred, preventing accurate reconstructions of imaged objects. This remains the case for each optical microscope that measures transmitted or reflected light, or a diffraction pattern, such that there are few standard metrics for images obtained \cite{paschali2019deep,perone2019promises}. Nonetheless, biological samples need to be imaged, identified, and characterized in detail and en-masse over their lifetime as various species pose a major threat to human health and safety.

Stratton \& Chu, in 1939, indicated the intensity of classical electric and magnetic fields cannot be rigorously described via the scalar Helmholtz equation assumed by Kirchoff's theory of diffraction \cite{stratton1939diffraction}. Instead, at least polarization or phase of the electric and magnetic fields at a detector are also necessary to reconstruct fields occurring in the image plane. Stratton \& Chu's electromagnetic diffraction theory motivated the design of interferometric microscopes, such as the confocal scanning microscope \cite{minsky1955confocal}, which achieves greater spatial resolution in the focal plane. However, these microscopes include guiding stars such as pinholes or slits which reject out-of-focus light and measure interference patterns as only partial solutions to electromagnetic diffraction theory.

Guiding stars in confocal microscopes, such as pinholes or slits, cause interference in the transmitted intensity such that the optical properties of the sample cannot be directly recovered from its measurement. Yet confocal laser scanning fluorescence microscopy is the standard microbiological assay, and involves applying labels or dyes to tag specific proteins in known cellular structures to identify anticipated cell species \cite{herman2020fluorescence}. Fluorophores adsorb to the surface of specific proteins,  providing local isotropic illumination and sharper image resolution. While direct, simple, and providing significantly greater spatial resolution than transillumination microscopy \cite{phillips2012measurement}, fluorescent dyes alter biological cells at the molecular level, are subject to photo-bleaching, and are incapable of tagging all different proteins or cell species of interest. Consequently, confocal microscopes are inherently biased to pre-selected information by the user. Nonetheless, laser-scanning confocal fluorescence microscopy continues to push new discoveries regarding biological cells and their structure.  Deep learning convolutional neural networks are further applied to compare fluorescent images with brightfield images,  attempting to automatically interpret imaging patterns for label free staining of the same biological sample \cite{ounkomol2018label,de2019deep,melanthota2022deep}.  Machine learning models for label-free staining or brightfield reconstruction require immense training datasets, which are typically not reproducible by smaller research teams. Alternative labelling methods with improved resolution include fluorescent light sheet microscopy, although this approach appears less common or standardized \cite{sapoznik2020versatile}. Machine learning was used to classify labeled microscope images of yeast cells belonging to different species, but yielded a relatively low accuracy of classification (74.6\%) given their similar appearance \cite{shankarnarayan2024machine}. Comparatively, light scattering classification has yielded greater accuracy for different species of biological cells \cite{sun2020deep}.

Angular-resolved dynamic light scattering can also be used to infer the size distribution and shapes of single cells from the far field \cite{burger1982extraction}. Compared with microscopic imaging, scattered light has the advantage of ensemble measurements of cell populations \cite{yang2004blood}. Light scattering patterns from yeast cells incubated after 1 day were classified with 98.3\% accuracy from yeast cells incubated after 4 days, suggesting a global time-dependence of a yeast cell's measured refractive index on the time of incubation \cite{eltigani2024light}. Changes in refractive index of the cell affect its image under a microscope, which requires greater quantification. Variation in refractive index of yeast cells has been reconstructed with less than 10 \si{nm} resolution using tomographic phase microscopy \cite{habaza2015tomographic}. Similar refractive index models of biological cells can be used to simulate light scattering in the design and analysis of label free cytometer systems.

Label-free whole cell reconstructions of refractive index have been attempted using a variety of other approaches including transillumination microscopy \cite{tong2024three}, optical diffraction coherence or phase tomography \cite{tan2006optical,choi2007tomographic,habaza2015tomographic, chen2020wolf}, X-ray microscopy\cite{merolle2023impact} or soft x-ray tomography\cite{mcdermott2009soft}, quantitative phase microscopy \cite{popescu2006diffraction,kujawinska2014problems,coppola2010digital}, holographic phase microscopy \cite{mico2008common,lee2011field,kemper2011simplified,rappaz2005measurement}, infrared vibrational spectroscopy \cite{wohlmeister2017differentiation}, Raman microscopy \cite{fernandez2023automatic}, and interferometry \cite{bon2009quadriwave,shaked2010reflective,shaked2012quantitative,girshovitz2012generalized,ding2010instantaneous}. However, each technique images from a limited numerical aperture, based on the Born-Rytov approximation (BRA) \cite{chen2020wolf}.  BRAs require small angle-light scattering measurements of optical phase \cite{tan2006optical,choi2007tomographic,habaza2015tomographic, chen2020wolf}, or polarization, over multiple angles.  Despite the promise of tomographic imaging and light scattering for reconstructing the refractive index of biological cells, few studies assess the accuracy of the BRA in heterogeneous media, despite its importance for inferring the physical properties of the media \cite{phillips2012measurement}. Since refractive index reconstruction is a reciprocal problem to light scattering, evaluating the accuracy of Born-Rytov approximate static light scattering solutions from the reconstructed refractive index is sufficient for this purpose.

In the near-field region of biological cells, complete three-dimensional frequency-domain and time-domain numerical methods including the discrete dipole approximation (DDA) \cite{draine1994discrete, yurkin2007discrete} and the Yee-lattice finite-difference time-domain (YL-FDTD) method \cite{liu20053, liu2023multi}, are theoretically more accurate than the BRA for recovering static light scattering patterns. Frequently, three-dimensional numerical methods with far-field projections (FFPs) are used to simulate the static light scattering patterns of biological cells. In turn, FFPs can be used in the design of new cell characterization or classification methods and systems \cite{wyatt1968differential, salzman1975flow, chen1982determination,singh2004analysis, sun2012speckle, liu2023multi}. Alternatively, the multi-slice Fourier transform (MSFT) method based on the BRA can also be used \cite{colombo2023three}. Each numerical method has advantages and limitations which are rarely compared for complex heterogeneous refractive index media \cite{yurkin2007discrete}. However, each numerical method shares a quantifiable absolute error compared with analytical Mie theory. In turn, the most accurate numerical approximation may be used to quantify the relative error of the less accurate numerical methods based on the static light scattering patterns from the complex heterogeneous refractive index reconstructions of biological cells. The relative error helps decide which method is sufficiently rigorous for the simulation of light scattering from biological cells and optical systems, given the time and accuracy required for their modeling. In turn, this informs which technique provides higher-accuracy light-scatter patterns for testing biological cell classifiers.

\section{Electromagnetic diffraction theory}
Kirchoff introduced scalar solutions for the propagation of light intensity within a closed domain according to the Helmholtz equation
\begin{equation}
\nabla^2\psi(\mathbf{x})+k^2\psi(\mathbf{x}) = 0,
\end{equation}
as the integral form
\begin{equation}
u(\mathbf{x}') = \frac{1}{4\pi}\int_S\left[\left(\frac{\partial \psi}{\partial n}\right)_S \frac{e^{ikr}}{r} - \psi_S\frac{\partial }{\partial n}\left(\frac{e^{ikr}}{r}\right)\right]ds
\end{equation}
where $\mathbf{x}$ is the position of the measured light intensity $\psi$, $\mathbf{x}'$ is the position of the predicted light intensity, $u$, within the bounding surface, $S$,  provided its wave-number, $k$, separation distance $r=||\mathbf{x}'-\mathbf{x}||$, and path $n$ normal to $S$ and oriented toward the inside of the closed domain. Stratton \& Chu indicate that the intensity $\psi$ and its derivatives at all interior points are unique for either the Dirichlet measurement $\psi_S$, or the Neumann measurement $(\partial \psi/\partial n)_S$. However the two measurements are mutually dependent whereas Kirchoff's formula allows for their independence. Consequently, any values of $\psi_S$ or $(\partial \psi/\partial n)_S$ may be selected for (2) and still satisfy (1), yet the predicted solutions $u$ and $\partial u/\partial n$ may be different from $\psi$ or $\partial \psi/\partial n$ on $S$ in general \cite{stratton1939diffraction}.

Stratton \& Chu present a general theory for electomagnetic diffraction based on Maxwell's equations \cite{stratton1939diffraction}, 
\begin{equation}
\begin{split} \mathbf{E}(\mathbf{x}') = -\frac{1}{i\omega\epsilon} \frac{1}{4\pi} \oint_C \nabla \psi \mathbf{H}_\mathrm{in} \cdot d\mathbf{s} + \oint_C \psi\mathbf{E}_\mathrm{in}\times d\mathbf{s} ... \\ ... - \int_{S_\mathrm{in}}\left(\mathbf{E}_\mathrm{in}\frac{\partial \psi}{\partial n} - \psi \frac{\partial \mathbf{E}_\mathrm{in}}{\partial n}\right) ds
\end{split}
\end{equation}
and 
\begin{equation}
\begin{split}
\mathbf{H}(\mathbf{x}') = -\frac{1}{i\omega\mu} \frac{1}{4\pi} \oint_C \nabla \psi \mathbf{E}_\mathrm{in} \cdot d\mathbf{s} + \oint_C \psi\mathbf{H}_\mathrm{in}\times d\mathbf{s} ... \\ ... - \int_{S_\mathrm{in}}\left(\mathbf{H}_\mathrm{in}\frac{\partial \psi}{\partial n} - \psi \frac{\partial \mathbf{H}_\mathrm{in}}{\partial n}\right) ds
\end{split}
\end{equation}

which also requires an estimate of the enclosed fields $\mathbf{E}_{\mathrm{in}}$ and $\mathbf{H}_{\mathrm{in}}$ at the boundary. Therefore, the fields and the image cannot be wholly reconstructed from light intensity alone, which requires additional measurements of polarization and phase of electromagnetic radiation reaching a detector. In order to bypass this constraint, (3) and (4) are modified using approximate forms. Such forms are typically based on the BRA \cite{habaza2015tomographic,chowdhury2019high}, orthographic projection \cite{wang2019simple}, or anomalous diffraction \cite{yang2004blood}. Alternatively, a boundary condition is assumed in order to estimate the phase and reconstruct the refractive index from variation of the intensity along the path of transmitted radiation \cite{phillips2012measurement}. The appropriateness of an approximate inverse solution depends on the region of interest, the shape and distance of the optics, the variation in the refractive index of the sample, and the intended purpose for the images obtained. Since each approximation is based on scattered light intensity, or its derivative, it is therefore subject to the same limitation as (2). 

 Likewise, complete analytical solutions for the inverse problem of solving the electric permittivity $\epsilon(\mathbf{x}')$, or magnetic permeability $\mu(\mathbf{x}')$ based on the time-reversal of far-field scattering patterns are available \cite{lai2022fast}, but have not been demonstrated for complicated heterogeneous refractive index distributions. In applications of flow cytometry, it is critical to assess the accuracy of the BRA for heterogeneous refractive index distributions.

Given the Mie analytical solution for electromagnetic scattering from a spherical geometry, the error of three-dimensional numerical methods for Maxwell's equations can be directly evaluated. Similarly, given numerical solutions for the complete Maxwell's equations, the relative error of less accurate BRAs for electromagnetic scattering from heterogeneous refractive index distributions may be inferred. In assuming the reciprocity of Maxwell's equations, the relative error of the light scattering pattern is synonymous with the relative error of the effective refractive index tomographic reconstruction, providing an estimate for its minimum spatial resolution.

\section{Analytical methodology}
Analytical solutions for electromagnetic scattering from spherical geometries can also be evaluated both in the far-field \cite{laven2003simulation}, and everywhere in the near field \cite{Zhu_2020} according to Mie theory. Scattering behaviour of arbitrary shapes has also been presented using T-matrix theory \cite{mishchenko2004t}. However, biological cells seldom have well-defined prismatic or periodic structures amenable to an exact analysis of Maxwell's equations based on symmetry arguments. Instead, numerical methods are required to evaluate the scattering of electromagnetic waves from arbitrary geometries or heterogeneous refractive index distributions. To the author's knowledge, complete numerical solutions for Maxwell's equations are seldom used to assess the accuracy of the the BRA for reconstructing a heterogeneous refractive index. However, such comparison is achievable given the known numerical approximation error with respect to a well-defined analytical solution on the same grid, which also allows for a point-to-point comparison of the near fields and far-field scattering patterns for an angular error metric.

\section{Numerical methodology}
The numerical methods compared include the ANSYS\textsuperscript{\textregistered} Lumerical finite-difference time-domain (FDTD) method, the DDSCAT 7.3.3 discrete dipole approximation (DDA), and multi-slice Fourier transform method referred to as \texttt{PyScatman} \cite{colombo2023three}. Linearly polarized time-harmonic optical radiation of 405 \si{nm} wavelength is imparted as a plane-wave along propagating in the positive $\hat{z}$ direction. The refractive index is given a uniform extinction coefficient ($\kappa=0.01$) such that the complex conjugate gradient (CCG) method in DDSCAT 7.3.3 may converge.

\subsection{Finite-difference time-domain method}
In this work, ANSYS\textsuperscript{\textregistered} Lumerical 2022 R2.4 is used as a popular commercial implementation of the finite-difference time-domain method (FDTD) method.
\subsubsection{Geometry and mesh refinement}
 In order to establish a benchmark, uniform lattice and non-uniform conformal mesh types are compared for the case of a sphere and at $N_\lambda=20$ units per wavelength. The most accurate automated mesh size distribution is used with a rendering of an S. \emph{cerevisiae} (Baker's yeast) cell from measurements in Ref. \cite{habaza2015tomographic}. To import the irregular geometric reconstruction of S. \emph{cerevisiae} into ANSYS\textsuperscript{\textregistered} Lumerical, it is first exported as stereolithographic triangulated mesh (STL) files representing isosurfacess of refractive index variations $\Delta n=0.01$ obtained in MATLAB R2023b. Small loss in the form of an extinction coefficient is added, $\kappa=0.01$ in order to compare its accuracy with the discrete dipole approximation (DDA) method.

\subsubsection{Source condition}
Plane wave excitation from a total-field scattered-field (TFSF) boundary is assumed using continuous wave normalization based on the  Broadband Fixed Angle Source Technique (BFAST). This is the default optimal method for discrete frequency transform (DFT) and far-field projection (FFP) in ANSYS\textsuperscript{\textregistered} Lumerical FDTD, even for the case of continuous wave simulation. 

\subsubsection{Far-field projection} The FFP is performed within the background medium, which is water with a refractive index varied slightly from Palik's measurements to neglect extinction ($n_b=1.33$). The advantage of ANSYS\textsuperscript{\textregistered} Lumerical FDTD is a FFP option for a dielectric interface prior to scattering, which is not intrinsic to other numerical methods. One major disadvantage of the FFP in ANSYS\textsuperscript{\textregistered} Lumerical FDTD, is that it is not performed automatically during run-time. This constrains the FFP to manual interactive-mode deployment that is unavailable in non-interactive high-performance computing (HPC). More recent versions of Lumerical may support the use of \texttt{*.lsf} analysis script files, but newer licenses were not available on the HPC nodes. Consequently, we were restricted to systems for which our RAM was sufficient to perform the DFT and FFP analysis. 

\subsubsection{Components of the far field} The FFP in ANSYS\textsuperscript{\textregistered} Lumerical FDTD, in addition to intensity, provides the electric field along each plane of incidence (polarized field components $E_p$ and $E_s$) and the subsequent phase which can be used directly within the Born and Rytov approximations for reconstructing the biological cell refractive index.

\subsubsection{Graphical acceleration} ANSYS\textsuperscript{\textregistered} Lumerical FDTD may be accelerated using a graphics processing unit (GPU) as of ANSYS\textsuperscript{\textregistered} Optics 2023, but this is only available for non-lossy media, and the total-field scattered-field (TF/SF) boundary condition required for the FFP is not supported by the GPU acceleration. Since GPU acceleration is not available for ANSYS\textsuperscript{\textregistered} Lumerical FDTD simulations, its comparison with DDA is based on CPU operations only.

\subsection{Discrete dipole approximation}
We employ the discrete dipole approximation (DDA) FORTRAN 90 code developed by Bruce T. Draine and Piotr J. Flatau, titled DDSCAT \cite{draine1994discrete}. We use the latest version DDSCAT 7.3.3 updated on January 20\textsuperscript{th} 2022 \cite{flatau2014light,draine_ddscat_website}. DDSCAT is implemented in a virtual environment, Windows Subsystem Linux (WSL), following the recommendations of the program authors.

\subsubsection{Geometry and mesh}
The geometry and properties of the sphere and S. \emph{cerevisiae} cell are implemented based on the interdipole spacing in DDSCAT 7.3.3. Since DDA is a finite element method based on  Green's function for point dipoles, a smaller grid can be used to achieve the same accuracy as the FDTD or MSFT methods. Instead of direct comparison, the rate of numerical error with respect to the units per wavelength is extrapolated based on an empirical fitting using the Levenberg-Marquardt algorithm \cite{levenberg1944method}. This allows an analytical solution representing exactly the same grid for each scattering analysis, and compares compute metrics for larger grids than are possible on a single compute node. DDSCAT requires a uniform voxel grid that is not amenable to mesh refinement, since its fast matrix inversion is based on complex conjugate gradient (CCG) or fast Fourier transform (FFT) method. This requires sufficient number of discrete dipoles to represent a smooth boundary, and further constrains the refractive index of the scatterer to have small variation compared with its background \cite{draine1994discrete}. In most microbiological applications, these approximation hold for effective refractive index reconstructions. However, ANSYS Lumerical FDTD is not limited to the same constraints.

\subsubsection{Source condition} Monochromatic plane waves are imparted with a 405 \si{nm} wavelength. In contrast to the BFAST method in ANSYS\textsuperscript{\textregistered} Lumerical FDTD, these wavelengths can be defined exactly since they are independent of the interdipole spacing, and instead appear directly within the analytical Green's function for each point dipole source. This avoids numerical dispersion from the source condition, but imposes lattice dispersion \cite{draine1994discrete} by assuming that Cardinal interdipole spacing is uniform with respect to the wavelength throughout the spatial domain.

\subsubsection{Far-field projection} By default, DDSCAT 7.3.3 performs a far-field projection (FFP) along one or more scattering planes during run-time execution. Scattering planes are defined for a single azimuthal angle $\phi$, or altitude angle $\theta$. These angles are defined with respect to the wave-vector or injection axis of the incident plane wave, as opposed to arbitrary Cartesian coordinates like in time-domain or frequency-domain solvers. The orientation of the scattering object is instead rotated with respect to the injection axis, and the differences in the frame accounted for by rotating the far-field pattern in $\theta$ and $\phi$ to the same axis for comparison. We extended the upper bound to the number of points for $\theta$ and $\phi$, \texttt{MXSCA=1000}, that can be used in DDSCAT 7.3.3, arbitrarily introduced by the authors to limit memory requirements, in order to represent the complete $4\pi$ \si{str} far-field scattering pattern with $\theta\in[0^\circ,179^\circ]$ and $\phi\in[0^\circ,359^\circ]$. Message passing interface (MPI) and OpenMP in DDSCAT 7.3.3 is still experimental, so DDSCAT 7.3.3 was implemented on a single logical processor for this study. Multiple instances of DDSCAT 7.3.3 were executed concurrently on the same CPU (13th Gen Intel\textsuperscript{\textregistered} Core\textsuperscript{TM} i7-13700K, 3400 Mhz, 16 Core(s), 24 Logical Processors).
\subsubsection{Components in the far field} Unlike the FDTD and MSFT methods, DDSCAT 7.3.3 directly returns up to 12 Mueller matrix scattering coefficients and the polarization of the projected far field, whereas Lumerical FDTD returns the intensity, $\mathrm{S}_{11}$, and electric field components $E_p$ and $E_s$. Conversions exist between the two, which allow us to draw further comparisons.

\subsubsection{Graphical acceleration} As far as the authors are aware, DDSCAT 7.3.3 does not include an option for graphical processing unit (GPU) acceleration, which it would greatly benefit from given its large matrix inversion using FFT. Although the technique has been implemented on high-performance computing (HPC) clusters and accelerated for GPU parallel processing \cite{flatau2014light,seeram2016acceleration}, these additional steps are not performed in this work given the single-node restricted comparison with ANSYS\textsuperscript{\textregistered} Lumerical FDTD.

\subsubsection{Parallelism}
Since the DDA is ``ridiculously parallelizable," DDSCAT 7.3.3 has been implemented with message passing interface (MPI) on multiple processors, each representing separate scattering angles (orientations of a scattering object). However, MPI has no advantage for a single scattering angle in DDSCAT 7.3.3, such that we restricted each simulation to one processor. Therefore, on a multi-core processor, multiple instances of DDSCAT 7.3.3 can be executed concurrently and independently, provided appropriate RAM and number of processors. To give fair evaluation of performance, this work draws comparisons between the numerical methods in terms of processor-hours.

\subsection{Multi-slice Fourier Transform} There are several implementations of the Born approximation available. One such method, which is valid in the small angle approximation, is \texttt{PyScatman} \cite{colombo2022scatman}. Despite its use primarily for x-ray scattering, it is applicable at longer wavelengths and larger scattering geometries, with an appropriate matching of the wavelength to the refractive index ratio. Implemented in CUDA\textsuperscript{\textregistered}, by default it requires GPU acceleration. It's run time is orders of magnitude less than the previous techniques, even without GPU acceleration, but its error requires further assessment. Given a small-angle approximation, \texttt{PyScatman} is restricted to a cone angle of 120$^\circ$. Its basis on a photon density at each layer restricts the forward scattering pattern to intensity alone, and is not amenable to a near-field analysis. Alternative Born approximations based on Maxwell's equations are not implemented in this work, except in the inverse case, since each operates on equivalent small-angle approximations.

\subsubsection{Geometry and mesh} Any arbitrary geometry may be imported into the solver, which slices the geometry according to the wavelength. The meshing used is uniform, and although it can be nonuniform, a uniform mesh is optimal given the FFT used.

\subsubsection{Material} The material imported requires a very small difference in refractive index from its background, $|\delta|<0.1$ and $|\beta|<0.1$, given
\begin{equation}
    n=1-\delta+i\beta.
\end{equation}
The reason for this constraint is that the small angle approximation used assumes 
\begin{equation}
    q_z=k_0\sqrt{1-\cos^2[\theta(q_x,q_y)]}\approx k_0(1-\cos[\theta(q_x,q_y)]) 
\end{equation}
based on the binomial approximation, where $\theta(q_x,q_y)=\cos^{-1}[1-(q_x^2+q_y^2)/k_0^2]$, $(x,y)$ are the transverse coordinates for the given longitudinal coordinate $z$, whereas $q_x$, $q_y$, and $q_z$ are the corresponding wave-numbers in \si{m^{\mbox{-}1}}. This approximation is used within the multi-slice Fourier transform
\begin{equation}
    \Psi(q_x,q_y) = \sum_{s_z=0}^{S_z} \exp(ik_0\left\{1-\cos\theta\right\}s_z\Delta z)\mathcal{F}[\rho_s(x,y)],
\end{equation}
where $k_0$ is the free space wavenumber, $s_z$ is the slice number along the propagation axis $z$, and $\mathcal{F}[\rho_s(x,y)]=\psi(q_x,q_y)$ is the two-dimensional Fourier transform of the photon density which is a function of $q_x$ and $q_y$ only
\begin{equation}
    \psi(q_x,q_y) = \sum_{s_y=0}^{S_y-1}\sum_{s_x=0}^{S_x-1}\exp\left[i\frac{s_x\Delta x}{S_x}q_x+\frac{s_y\Delta y}{S_y}q_y\right] \Delta x \Delta y.
\end{equation}
In (5), $s_x$ and $s_y$ represent the slice index along $x$ and $y$ coordinates, whereas $S_x$ and $S_y$ are the total number of slices from these coordinates given grid spacing $\Delta x$ and $\Delta y$. Since the default background medium is that of free space $n=1.00$, the wavelengths are adjusted to return the same wave-number as for the background medium ($n=1.33$) in the far-field projection (FFP). This requires the material refractive index range is similarly truncated to $|\delta|<0.1$. This is accomplished via a proportional scaling of the differential refractive index $\Delta n$ distributed about the background refractive index $n_b$, such that $|\delta|=\Delta n/n_b$.

\subsubsection{Source condition}
The source condition is a monochromatic plane wave of $\lambda_s=405/n_b$ \si{nm} wavelength adjusted to the wave-number of the background medium used in the other numerical methods. Since $n_b=1.33$, the corresponding wavelength initialized in \texttt{PyScatman} is only $\lambda_s=342.1$ \si{nm} in order to match the scattering pattern from Mie theory.

\subsubsection{Far-field projection} The FFP is the only result returned by \texttt{PyScatman}, but to represent the type of scattering measurement performed. Only the ``\texttt{angular}" projection is compared.

\subsubsection{Graphical acceleration} Since CUDA\textsuperscript{\textregistered} is the default, this is the only method where GPU acceleration is used. In simulations performed using\texttt{PyScatman}, an Nvidia Quadro\textsuperscript{\textregistered} A4500 GPU is accessed and obtains scattering patterns in orders of magnitude less time, but with less accuracy.

\section{Cell Modeling Methodology}
Tomographic effective refractive index reconstructions of biological cells, such as S. \emph{cerevisiae}, are reported to nanometer resolution, despite the 405 \si{nm} wavelength selected. The 405 \si{nm} wavelength is selected since it represents the shortest wavelength in laser scanning confocal microscopes and visible flow cytometers.

 We selected the tomographic reconstruction of S. \emph{cerevisiae} by Habaza \emph{et al.} for our evaluation \cite{habaza2015tomographic}, given the optical tweezer used for its measurement, which provides high fidelity accurate static light scattering.

\subsection{Recovering the refractive index}
The refractive index is presented as an image stack video which we import to MATLAB\textsuperscript{\textregistered} R2023b using functions from the image processing toolbox. Afterwards, the cell is down-sampled to a feasible grid size for numerical computation, based on volumetric arithmetic mean averaging. Further information regarding the program for extracting the refractive index data is provided as Supplementary material.

\begin{figure}[H]
    \centering
    \includegraphics[width=0.8\columnwidth]{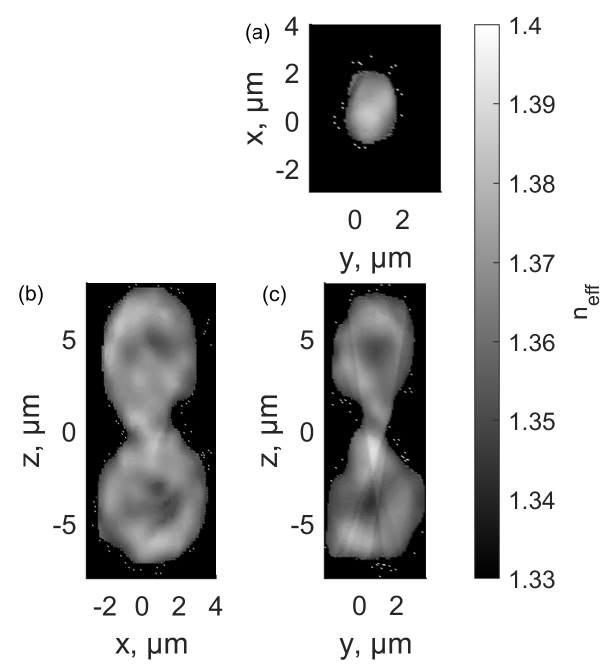}
    \caption{Empirically measured effective refractive index, which ranges from $n_b=1.33$ to $n_{\mathrm{nuc.}}=1.39$\cite{habaza2015tomographic}.  (a) The $xy$ cross section, (b) the $xz$ cross section, and (c) the $yz$ cross section. In general, the shape of refractive index regions in the cell are non-spherical and cannot be approximated by Mie scattering theory. In (c), sharp edges in the gray scale color map appear as a result of the Rytov approximation. This illustrated reconstruction is exact from measurements by Habaza \emph{et al.}\cite{habaza2015tomographic}.}
    \label{fig:1}
\end{figure}

Fig. \ref{fig:1} illustrates cross sections of the reconstructed S. \emph{cerevisiae} refractive index in MATLAB R2023b. The refractive index is heterogeneous, varying at nanometer scale, and is generally asymmetric and non-spherical. The cell scatters light in a manner that is challenging to solve analytically, but is amenable to numerical analysis. The BRA is used to reconstruct the S. \emph{cerevisiae} cell based on measured phase of scattered light \cite{habaza2015tomographic}. However, from the first stitching, it is apparent the BRA includes error that manifests in the $yz$ plane, exhibited by diagonal lines representing sharp step-functions or differences in refractive index. These step-functions occur despite fixation of the yeast cell in an optical trap \cite{habaza2015tomographic}. Such sharp differences suggest a lower spatial resolution of the measurement, along the propagation axis, than is presented by the three-dimensional reconstruction. We avoid assessing the accuracy of the measured reconstruction, and instead assess the accuracy of the BRA for electromagnetic scattering from a geometrically similar refractive index reconstruction truncated to eight distinct refractive index values. These adjustments are made to satisfy memory and compute constraints for the numerical methods employed. Therefore, the cell is down-scaled by a factor of 1.6 in each dimension to satisfy memory requirements on the computer used for simulation. The selected wavelength is also adjusted to 405 \si{nm} instead of the combined 485/532/569/1034 \si{nm} wavelengths used for its measurement, and the refractive index values could be different at this wavelength. Field simulations are not exactly representative of the experiment, but used for comparison between the numerical methods on a similar structure. 

\subsection{Selected numerical methods}
The discrete dipole approximation (DDA) is simulated using DDSCAT 7.3.3 \cite{draine1994discrete}, whereas the Yee-Lattice finite-difference time-domain method (YL FDTDM) is simulated using  ANSYS\textsuperscript{\textregistered} Lumerical FDTD 2022, and the Born-Rytov approximation (BRA) is implemented using the multi-slice fourier transform (MSFT) in \texttt{PyScatman} \cite{colombo2023three}.

\begin{figure}[H]
    \includegraphics[width=\columnwidth]{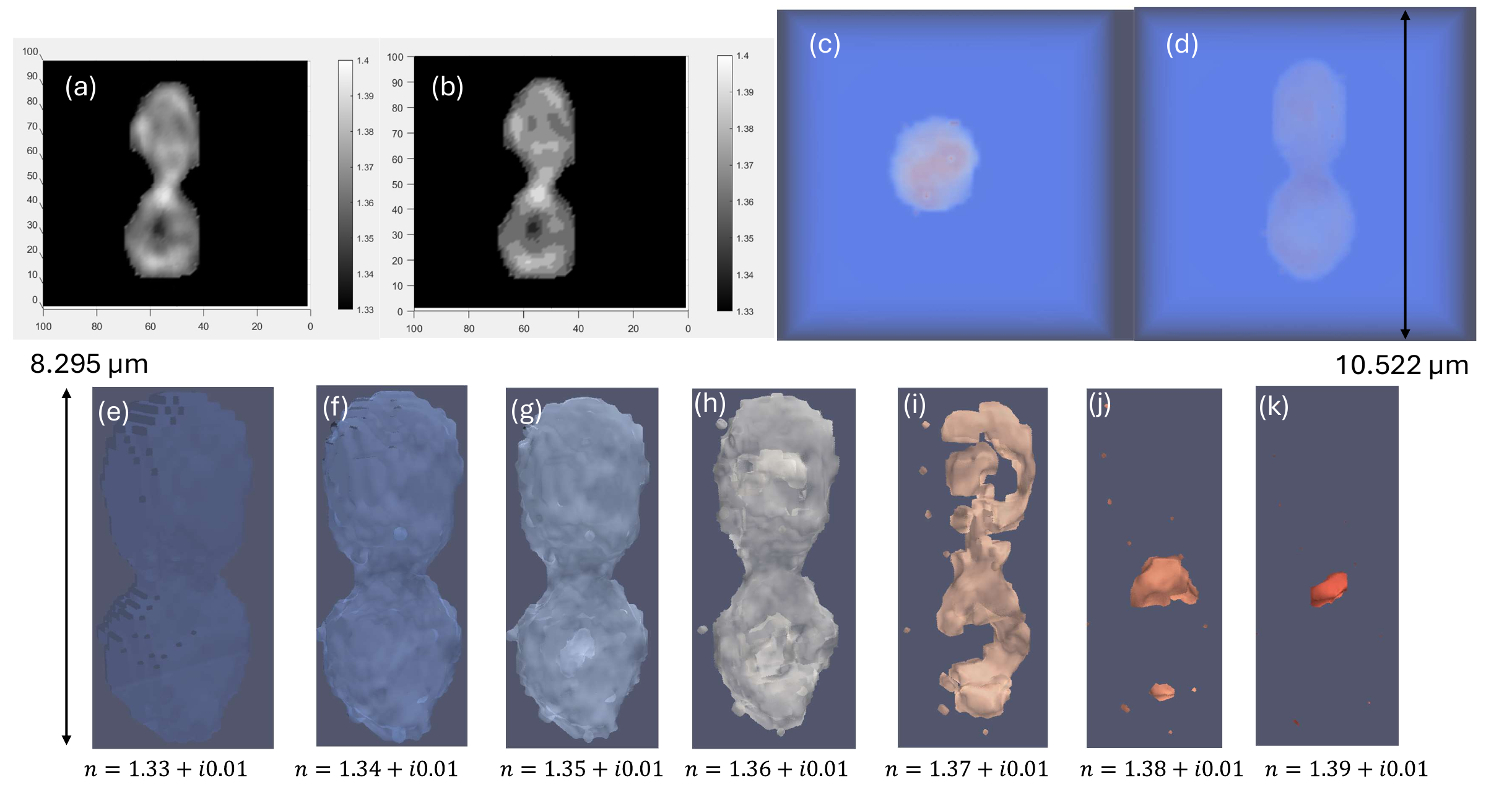}
    \caption{Illustration of cell partitioning. (a) Original cell refractive index down-sampled to 100 lattice units in each dimension, (b) mapped refractive index to seven discrete values. (c) and (d) are the top and side views of the cell, respectively, as Gaussian point distributions plotted in Paraview 5.12.0-RC2. (e-k) Gradations of $\mathcal{R}\{\Delta n\}=0.01$ in refractive index value from $n=1.33+i0.01$ to $n=1.39+i0.01$ plotted in Paraview 5.12.0-RC2 as a volume. The small extinction coefficient $\kappa=0.01$ is necessary for convergence of the solution within the empirically derived error tolerance. The same injection axis, cell dimensions, and orientation are used for direct comparison with ANSYS\textsuperscript{\textregistered} Lumerical FDTD (Fig \ref{fig:3}). These cell dimensions are a factor of 0.625$\times$ lesser than measured by Habaza \emph{et al.}\cite{habaza2015tomographic}.}
    \label{fig:2}
\end{figure}

\subsection{S. \emph{cerevisiae} in DDA}
The discrete dipole approximation (DDA) is simulated using DDSCAT 7.3.3 \cite{draine1994discrete}. Fig. \ref{fig:1} shows a mapping of the cell in \texttt{ParaView} using a custom \texttt{shape.dat} file which defines a matrix of lattice locations and their corresponding local composition. Each composition requires a separate refractive index file to be defined, similar to the case of ANSYS\textsuperscript{\textregistered} Lumerical FDTD, such that we define eight separate compositions varying with $\Delta n=0.01$, rounded to the nearest refractive index. Refer to Supplementary documents for an example.

\subsection{S. \emph{cerevisiae} in FDTD}
\begin{figure}[H]
    \centering
    \includegraphics[width=\columnwidth]{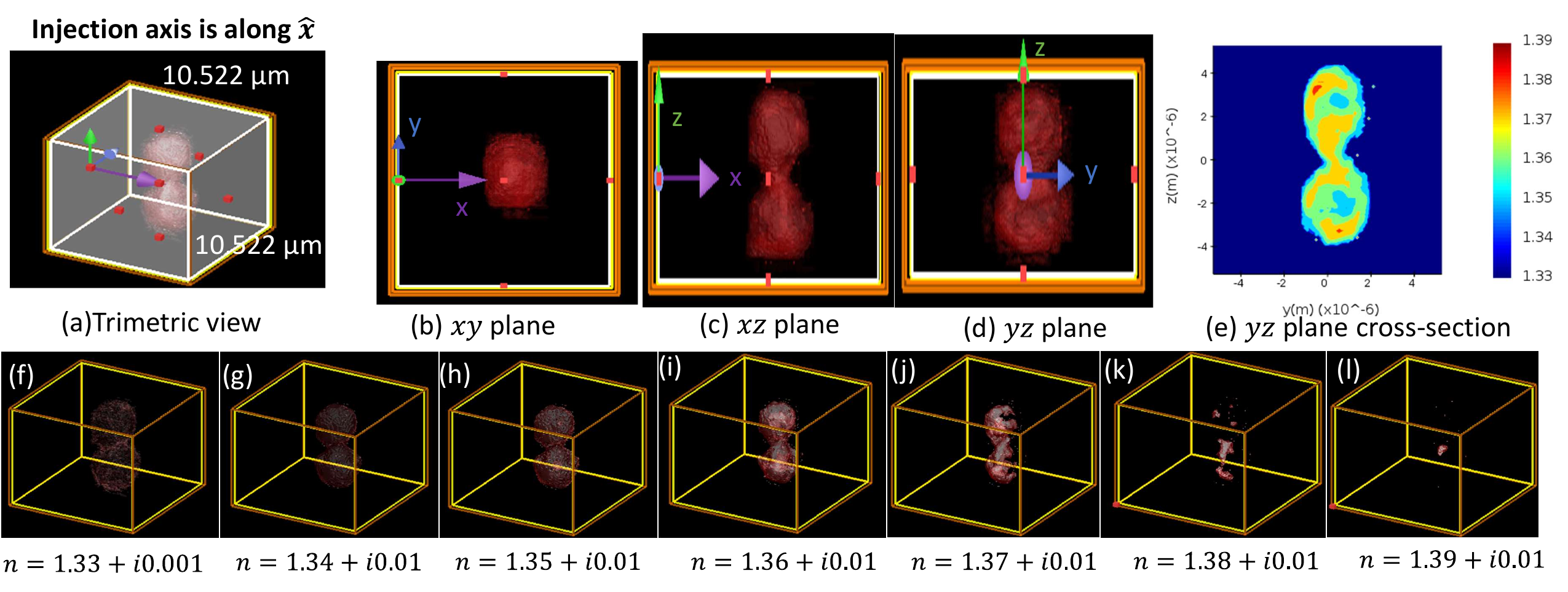}
    \caption{Tomographic model of S. \emph{cerevisiae} modeled in ANSYS\textsuperscript{\textregistered} Lumerical FDTD using seven discrete refractive index values. Each refractive index value is given a small extinction coefficient, $\kappa=0.01$, to compare with DDSCAT 7.3.3, which cannot perform its analysis on a completely lossless system. (a) The trimetric view of the geometry with increasing transparency mapped to refractive index. (b-c) side views of the same geometry in (a). (e) Refractive index of the cell returned from an \texttt{index monitor}. (f-l) Gradations of $\mathcal{R}\{\Delta n\}=0.01$ in refractive index value from $n=1.33+i0.01$ to $n=1.39+i0.01$. These cell dimensions are a factor of 0.625$\times$ lesser than measured by Habaza \emph{et al.}\cite{habaza2015tomographic}, and identical to Fig. \ref{fig:2}.}
    \label{fig:3}
\end{figure}

Fig. \ref{fig:3} shows the same structure of S. \emph{cerevisiae} reconstructed in ANSYS\textsuperscript{\textregistered} Lumerical FDTD. ANSYS\textsuperscript{\textregistered} Lumerical FDTD allows the import of STL mesh files which must conform to $(n,\kappa)$ materials pre-defined by the user. We partition the cell into eight refractive index spatial regions varying by $\Delta n=0.01$, using nearest refractive index rounding as a plotted \texttt{isosurface} in MATLAB R2023b, subsequently triangulated as an STL mesh. The effective refractive index generally increases toward the inside of the cell, such that the higher refractive index media are embedded in the lower refractive index media determined by Lumerical's \texttt{object heirarchy menu}. In general, this is not the case, so the embedding must be performed judiciously. This can become tedious for more heterogeneous cell morphologies. Notably, the reconstruction is an effective refractive index as opposed to the physical refractive index due to internal reflections that may occur within the cell which cannot be recovered from measurements of normal transmission. The STL mesh is forced to conform to the Yee-lattice by selecting a uniform grid with an integer spacing of the reconstruction. The more heterogeneous a cell refractive index is measured, the more ambiguous the connectivity of the STL mesh is defined. STL files must have complete connectivity for support in ANSYS\textsuperscript{\textregistered} Lumerical FDTD, otherwise they cannot be imported. Larger more heterogeneous cells are less likely to satisfy this requirement, which challenges their representation in ANSYS\textsuperscript{\textregistered} Lumerical FDTD.

\subsection{S. \emph{cerevisiae} in BRA}
The BRA solver, \texttt{PyScatman}\cite{colombo2023three} accepts geometric imports as a discrete array of refractive index nodes similar to the reconstruction in MATLAB 2023b and DDSCAT 7.3.3. Fig. \ref{fig:1} shows an illustration of this refractive index reconstruction. 

\begin{figure*}
    \centering
    \includegraphics[width=0.3\textwidth]{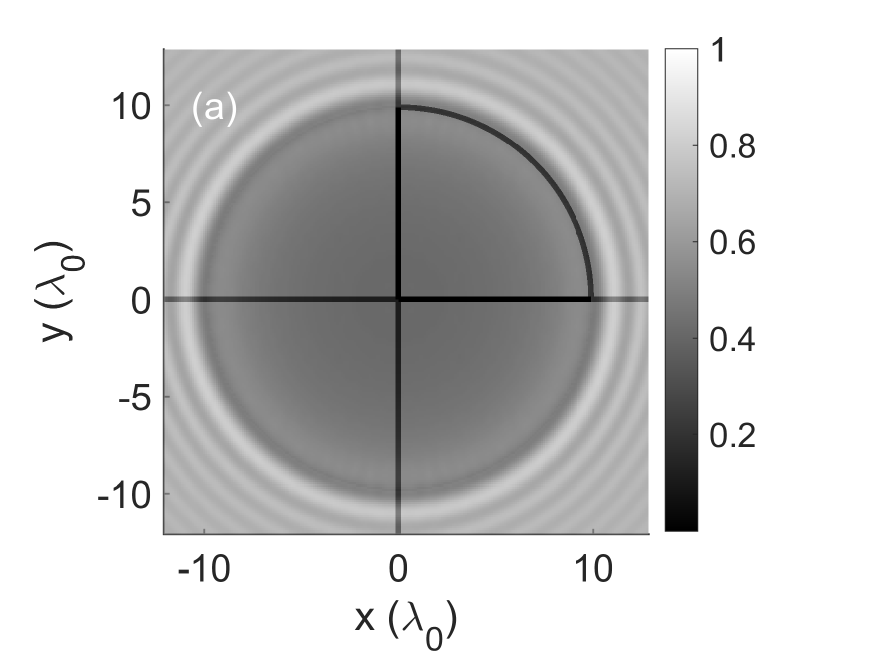}
    \includegraphics[width=0.3\textwidth]{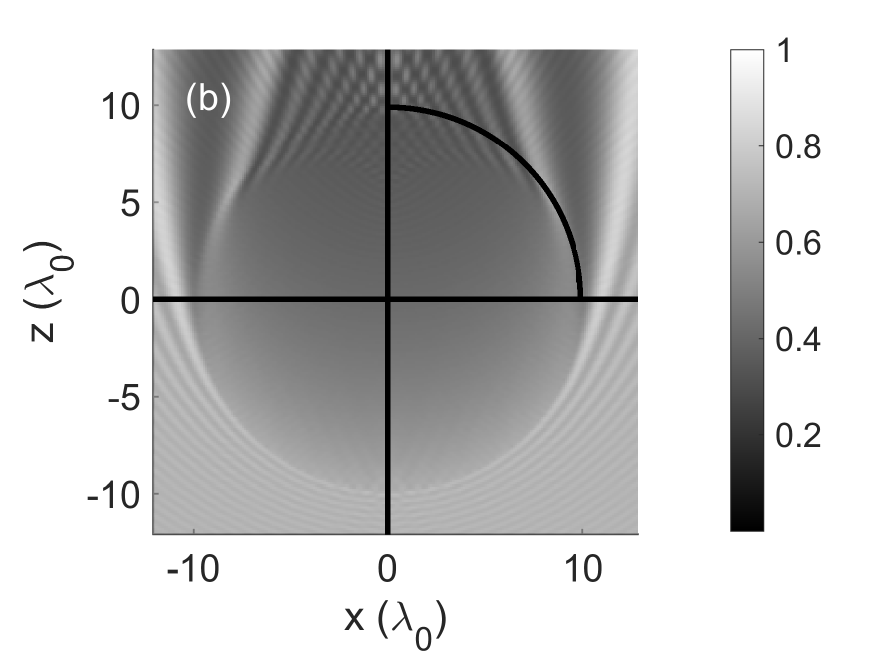}
    \includegraphics[width=0.3\textwidth]{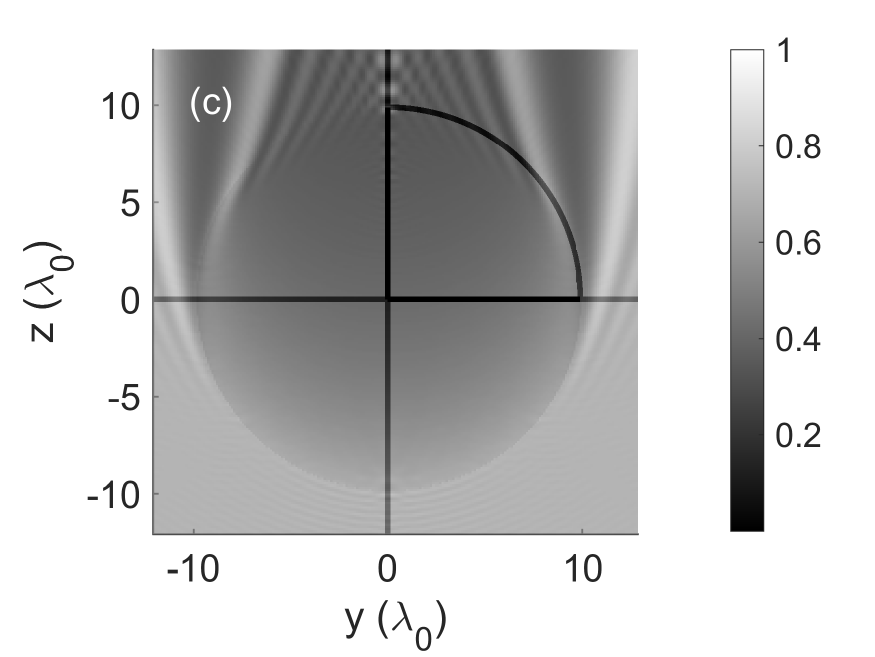} \\
    \includegraphics[width=0.3\textwidth]{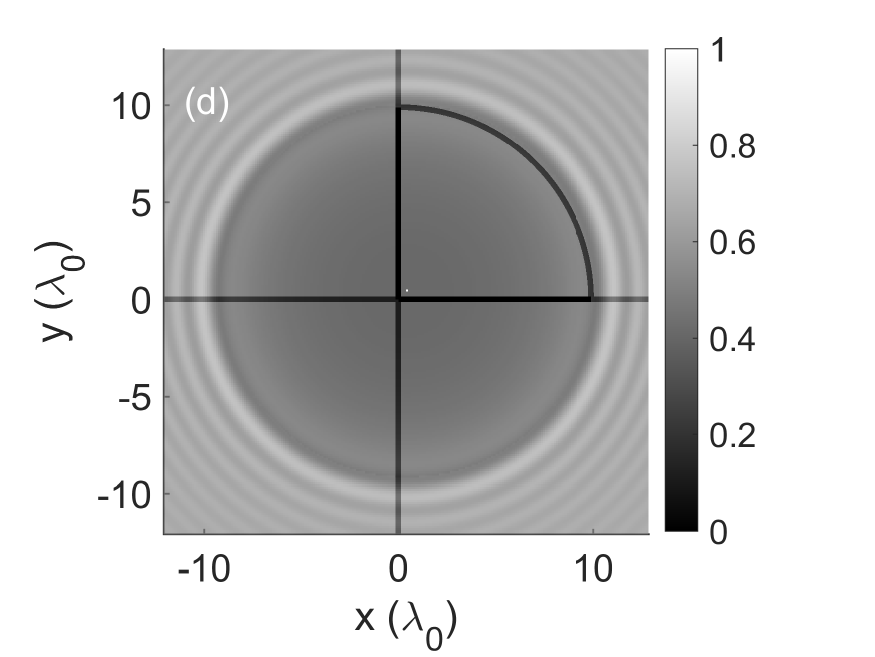}
    \includegraphics[width=0.3\textwidth]{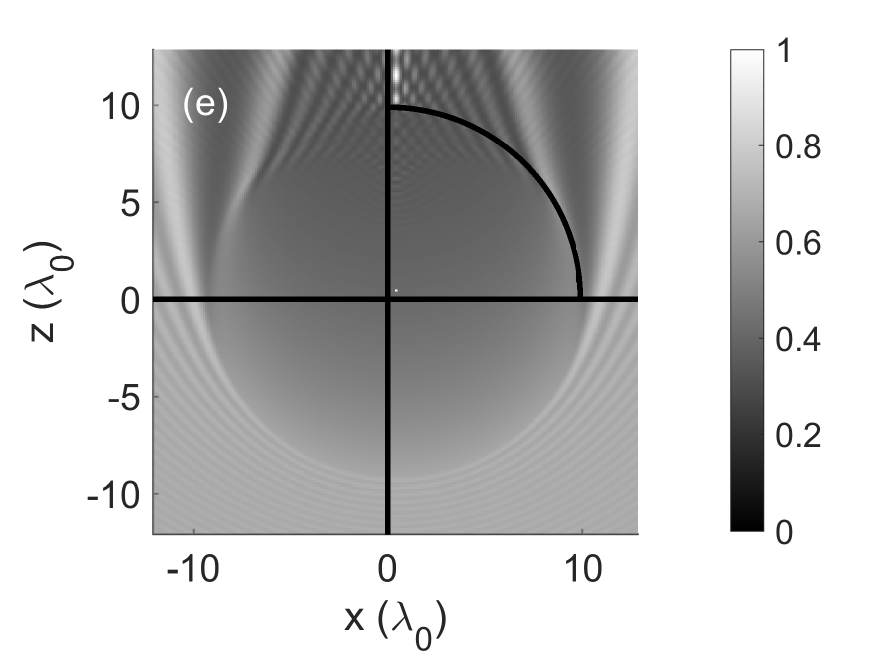}
    \includegraphics[width=0.3\textwidth]{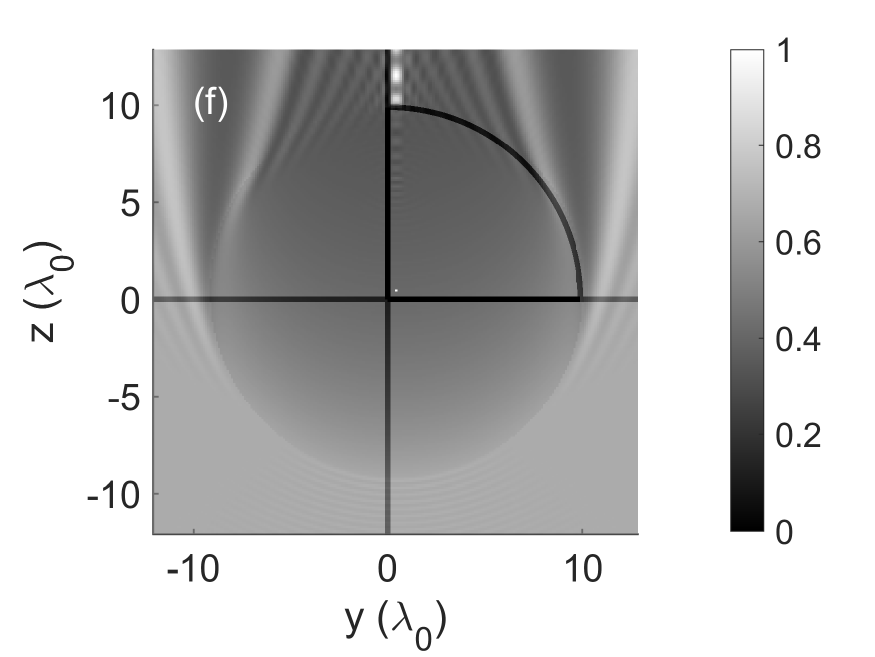} \\
    \caption{Near-field intensity ($\mathrm{S}_{11}$) predicted by (a, b, c) the discrete dipole approximation (DDA, DDSCAT 7.3.3\cite{draine1994discrete}) and (e, f, g) Mie theory\cite{Zhu_2020} for the case of $N_\lambda\approx9$ along three different planes of incidence: (a, d) $xy$, (b, e) $xz$, and (c, f) $yz$. Each plot axis is in terms of free-space wavelength, $\lambda_0$. }
    \label{fig:4}
\end{figure*}

\begin{figure*}
    \includegraphics[width=0.3\textwidth]{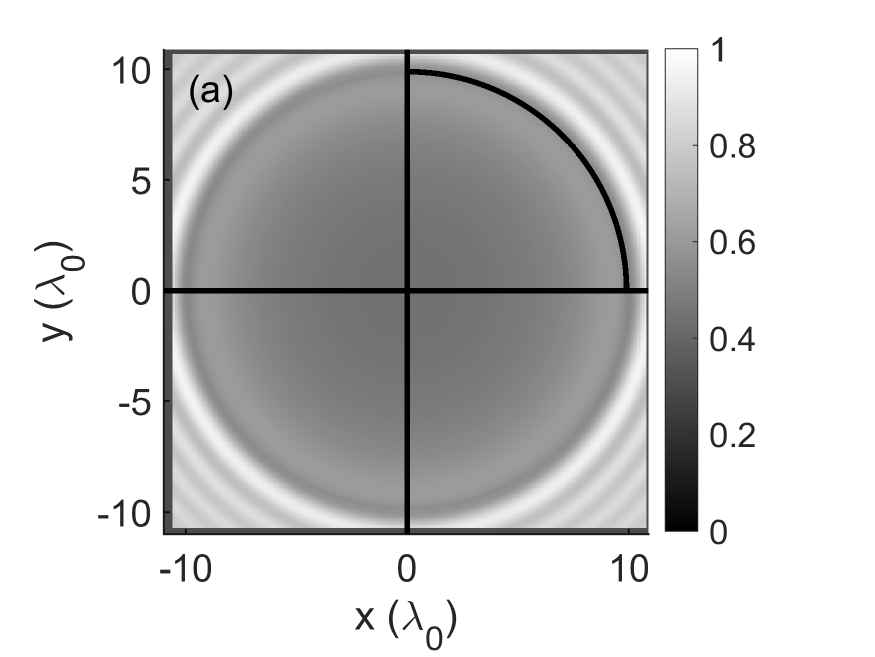}
    \includegraphics[width=0.3\textwidth]{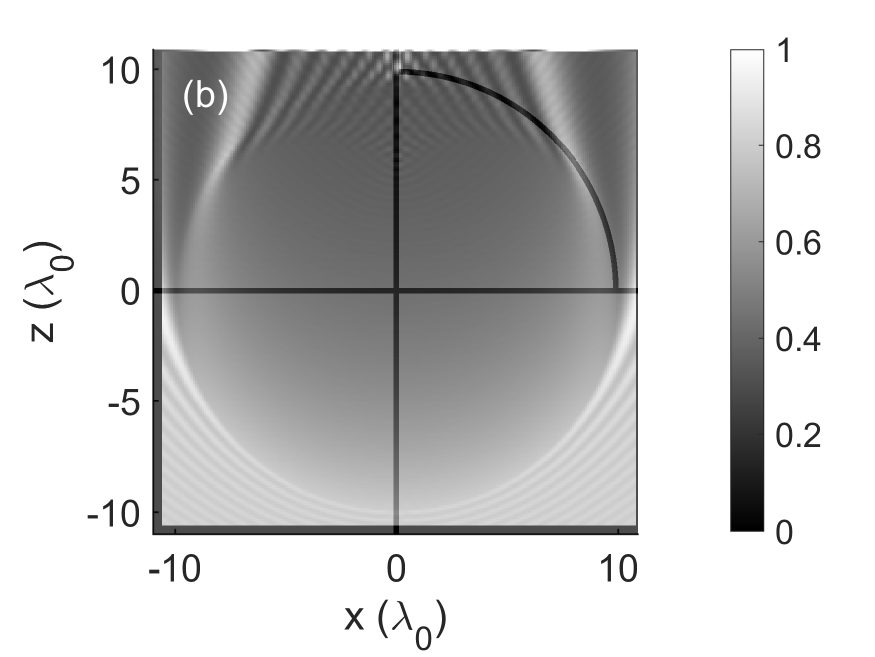}
    \includegraphics[width=0.3\textwidth]{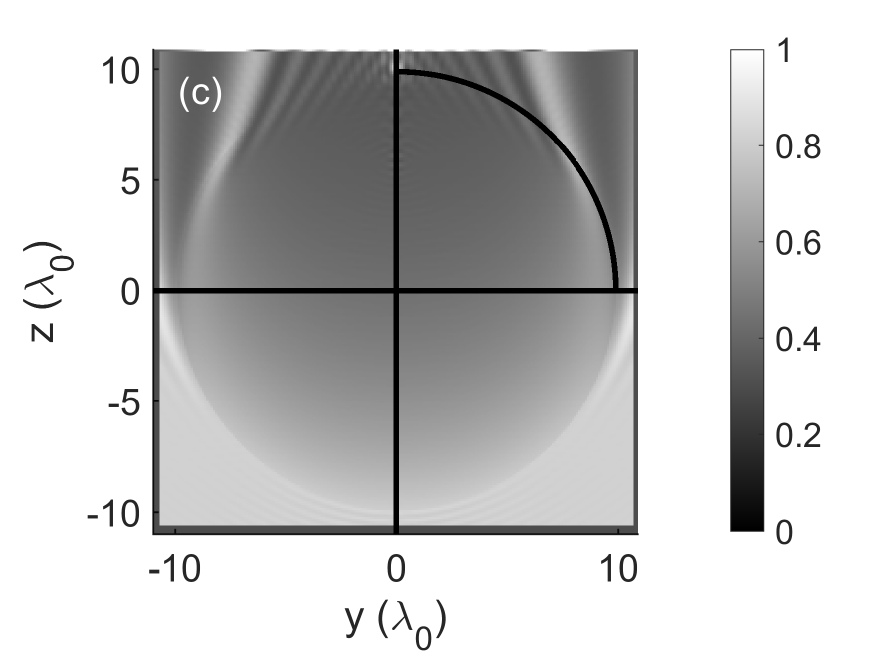} \\
    \includegraphics[width=0.3\textwidth]{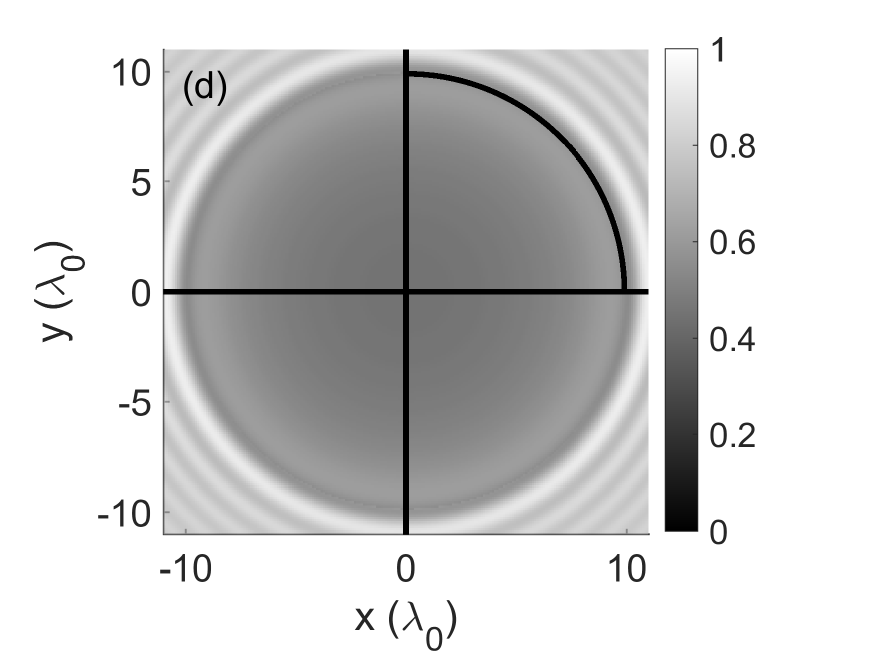}
    \includegraphics[width=0.3\textwidth]{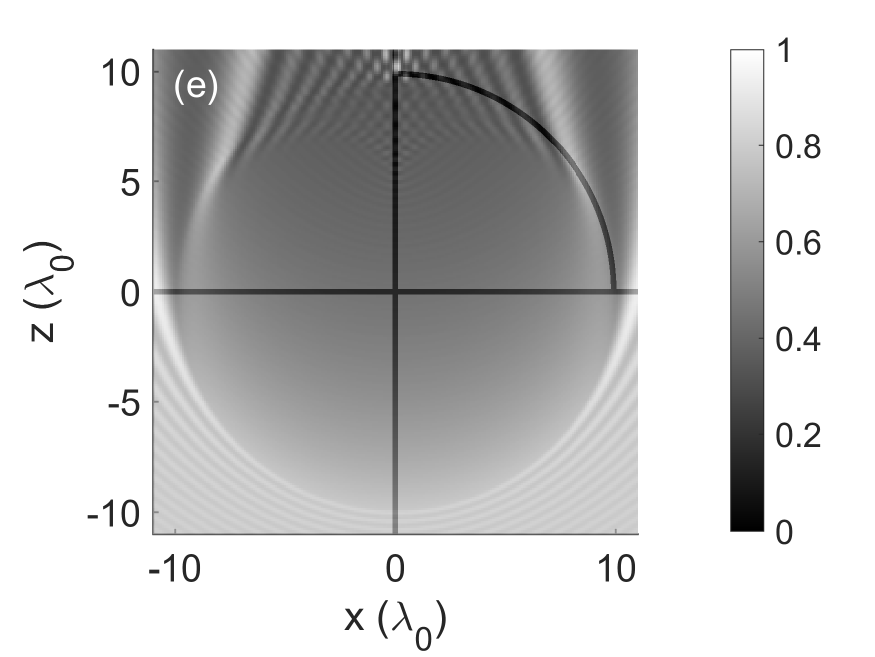}
    \includegraphics[width=0.3\textwidth]{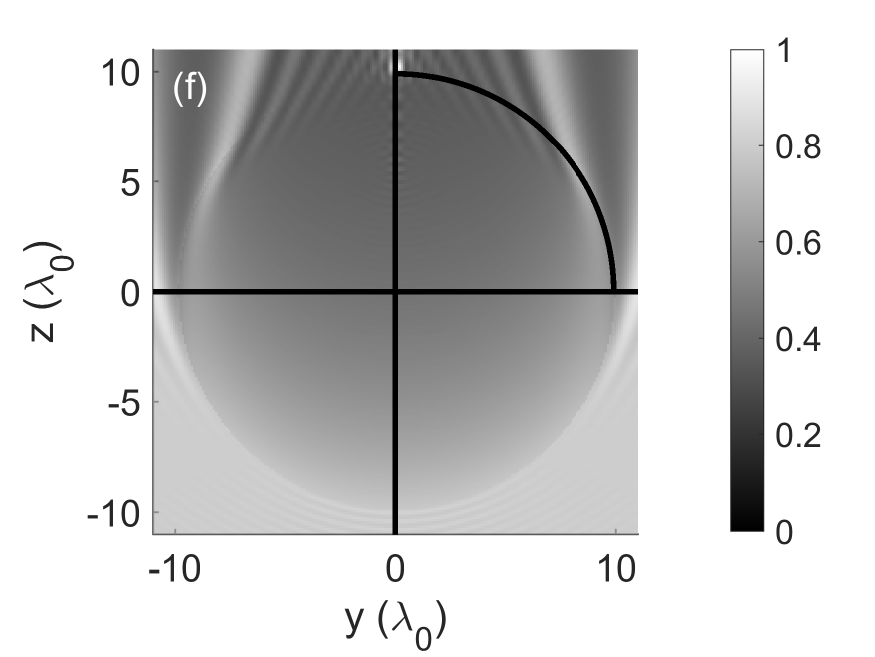} \\
    \caption{Near-field intensity ($\mathrm{S}_{11}$) predicted by (a, b, c) ANSYS\textsuperscript{\textregistered} Lumerical FDTD and (e, f, g) Mie theory\cite{Zhu_2020} for the case of $N_\lambda\approx11$ along three different planes of incidence: (a, d) $xy$, (b, e) $xz$, and (c, f) $yz$. Each plot axis is in terms of free-space wavelength, $\lambda_0$. }
    \label{fig:5}
\end{figure*}

\begin{figure}
    \centering
    \includegraphics[width=0.45\columnwidth]{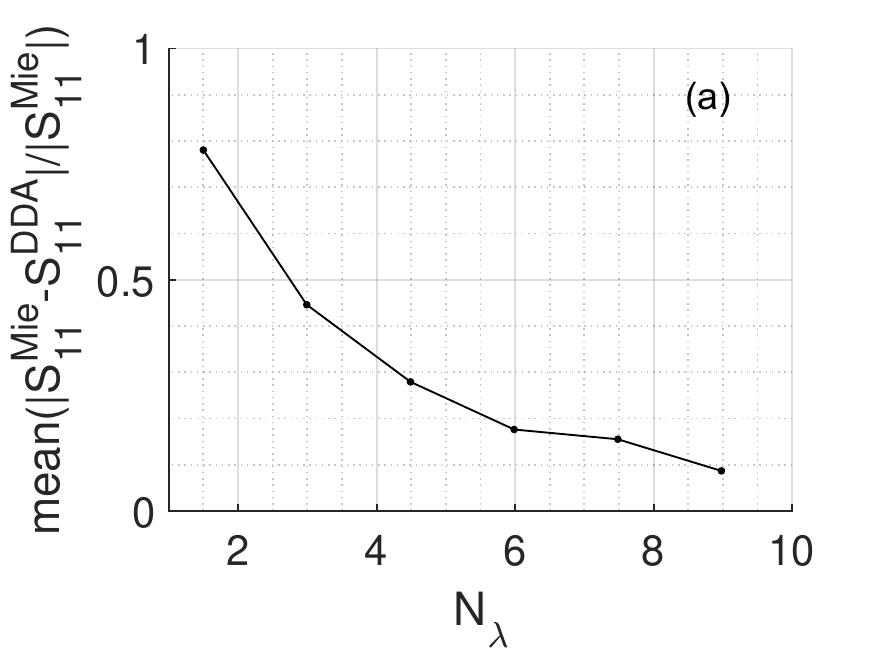}
    \includegraphics[width=0.45\columnwidth]{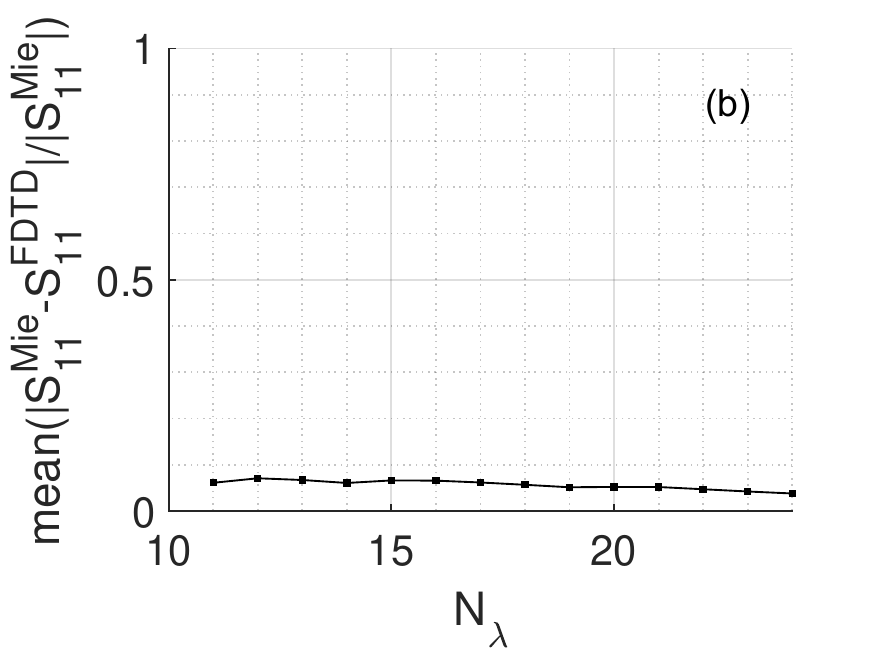}
    \caption{The mean error of the near-field in (a) DDSCAT 7.3.3 compared with (b) ANSYS\textsuperscript{\textregistered} Lumerical FDTD with respect to the units per wavelength, $N_\lambda$.  DDSCAT 7.3.3 has a similar trend and magnitude of error as ANSYS\textsuperscript{\textregistered} Lumerical FDTD, suggesting that the error of the DDA and FDTD methods are similar in the near-field. The advantage of the DDA is that, despite similar error, stability is maintained for lower $N_\lambda$. The advantage of FDTD is that a smaller near-field error can be achieved on the same compute node given lower memory requirements for the same number of grid units. }
    \label{fig:6}
\end{figure}


\begin{figure*}
    \includegraphics[width=\textwidth]{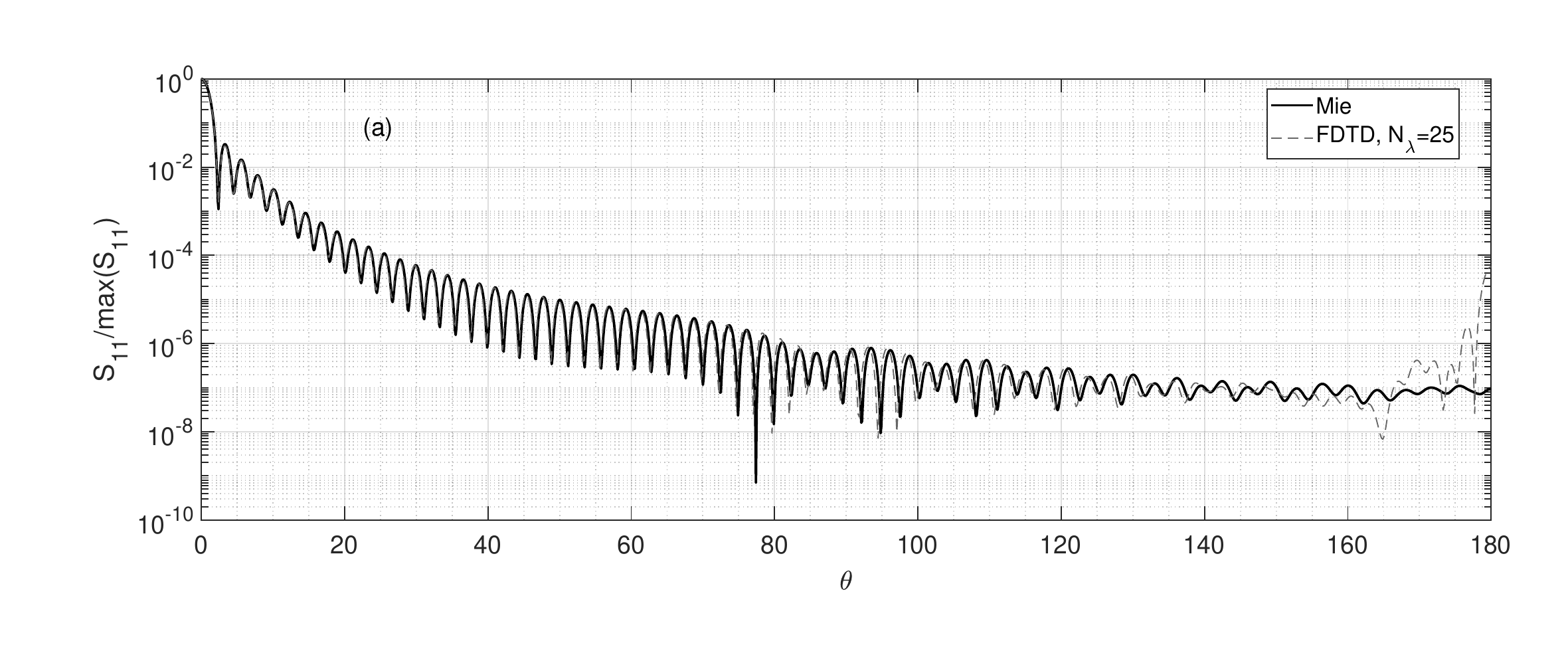}
    \includegraphics[width=\textwidth]{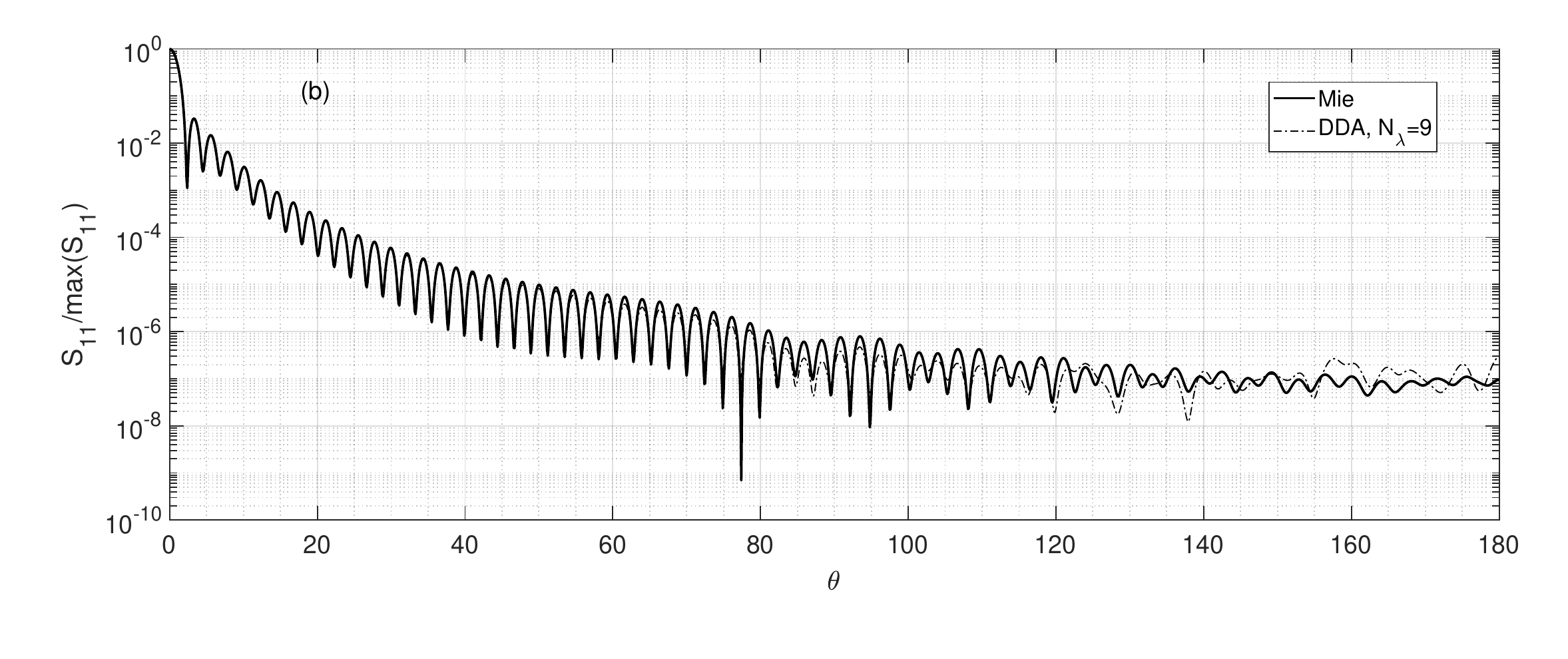}
    \caption{Mie theory compared with (a) ANSYS\textsuperscript{\textregistered} Lumerical FDTD at $N_\lambda=24$ (b) DDSCAT 7.3.3 at $N_\lambda=9$. Although each solution is in agreement with Mie theory, the FDTD method in (a) exhibits increasing angular deviation from Mie theory as $\theta$ increases. In contrast, the DDA method in (b) exhibits increasing magnitude deviation from Mie theory as $\theta$ increases. Ultimately, the angular error of FDTD is greater than the amplitude error of the DDA, but may be of sufficient merit for pattern recognition. }
    \label{fig:7}
\end{figure*}

\section{Results and Discussion}
First the DDA, YL-FDTD method, and the BRA MSFT method are compared with analytical Mie theory spherical scattering coefficient for intensity, $\mathrm{S}_{11}$ both in the near-field (Figs. \ref{fig:4}-\ref{fig:5}), and from a far-field projection (Fig. \ref{fig:7}). Second, a comparison of the scattering pattern is drawn for the case of the S. \emph{cerevisiae} cell. In each example, the error of the BRA can be inferred directly by comparison between the numerically evaluated far-field $S_{11}$ patterns (Fig. \ref{fig:8}). Absolute error
\begin{equation}
    \mathrm{err}=\frac{|\mathrm{S}^{\mathrm{Num}}_{11}-\mathrm{S}^{\mathrm{Mie}}_{11}|}{\mathrm{S}^{\mathrm{Mie}}_{11}},
\end{equation}
where $\mathrm{S}^{\mathrm{Mie}}_{11}$ and $\mathrm{S}^{\mathrm{Num}}_{11}$ are the analytically and numerically evaluated scattering coefficients associated with intensity. 
\subsection{Spherical scattering}

\begin{figure*}
    \includegraphics[width=0.3\textwidth]{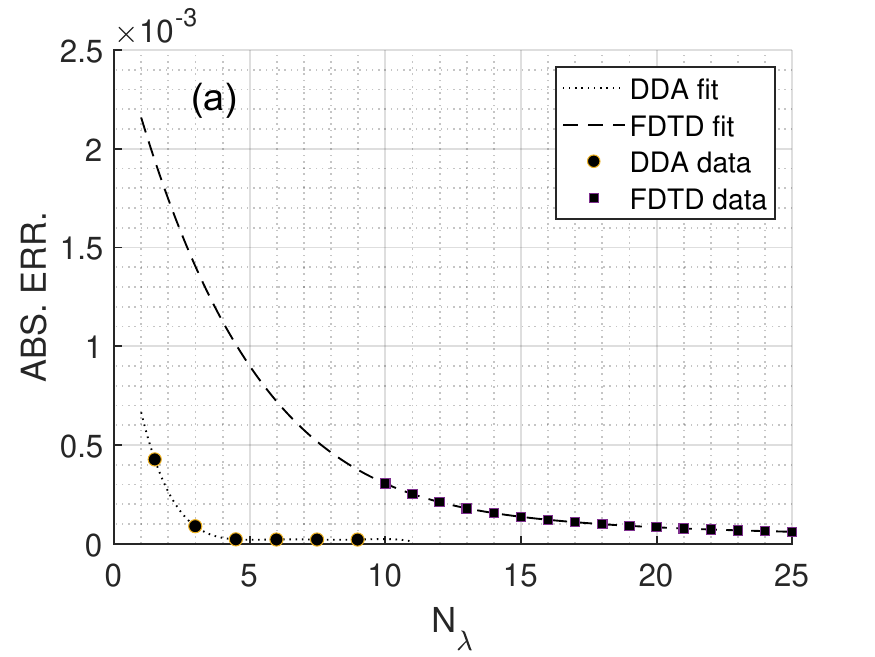}
    \includegraphics[width=0.3\textwidth]{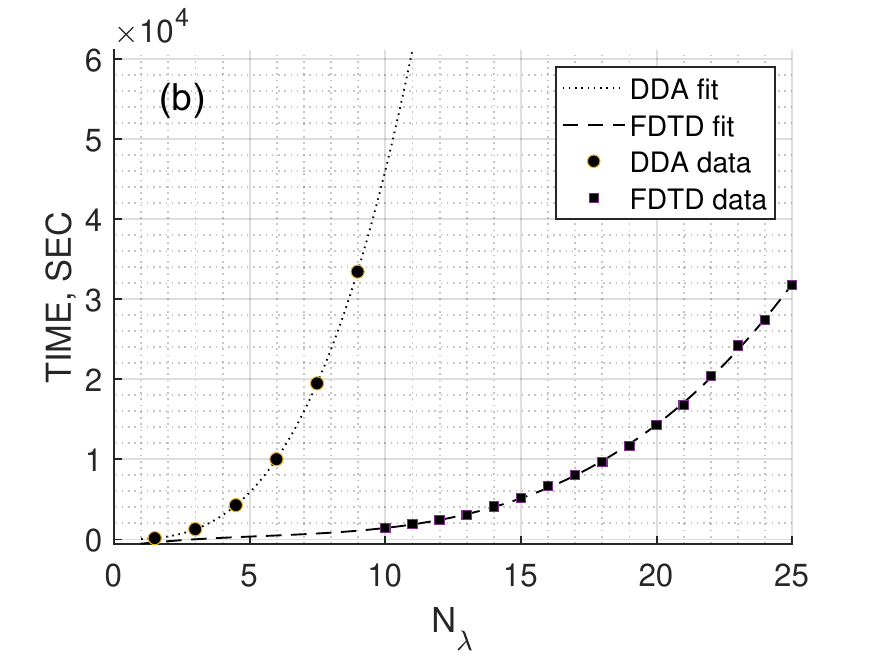}\includegraphics[width=0.3\textwidth]{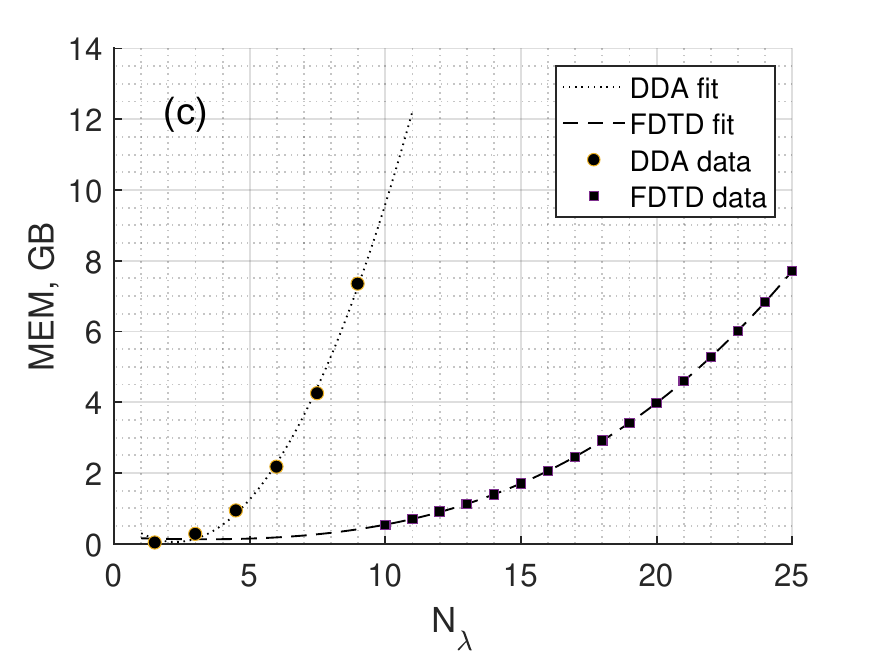} \\
    \includegraphics[width=0.3\textwidth]{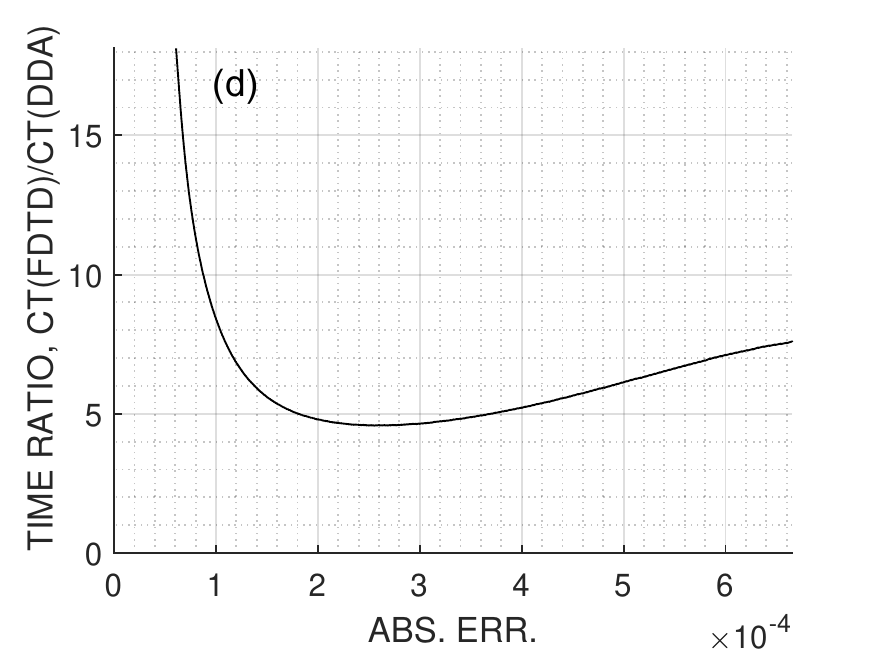}
    \includegraphics[width=0.3\textwidth]{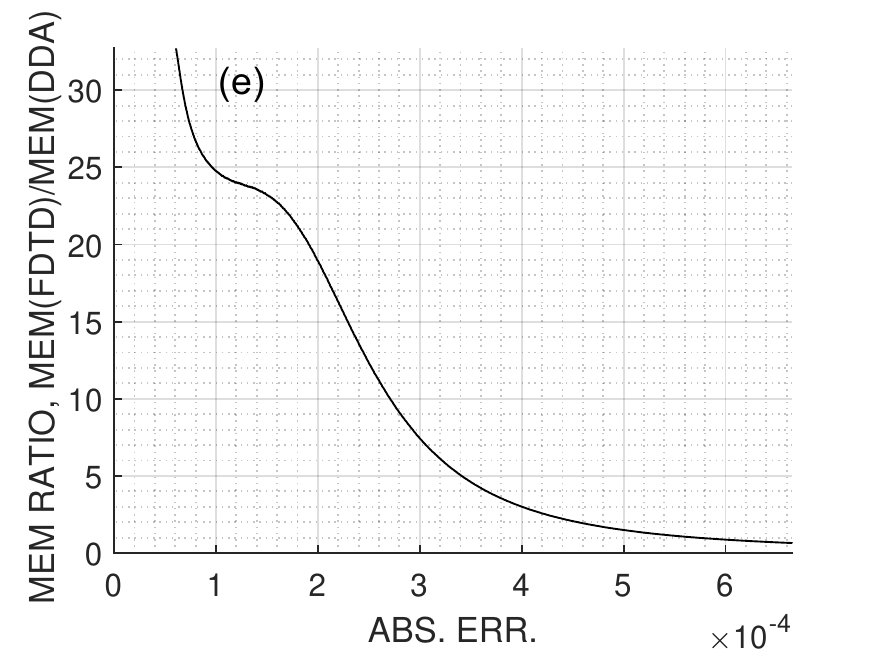}\includegraphics[width=0.3\textwidth]{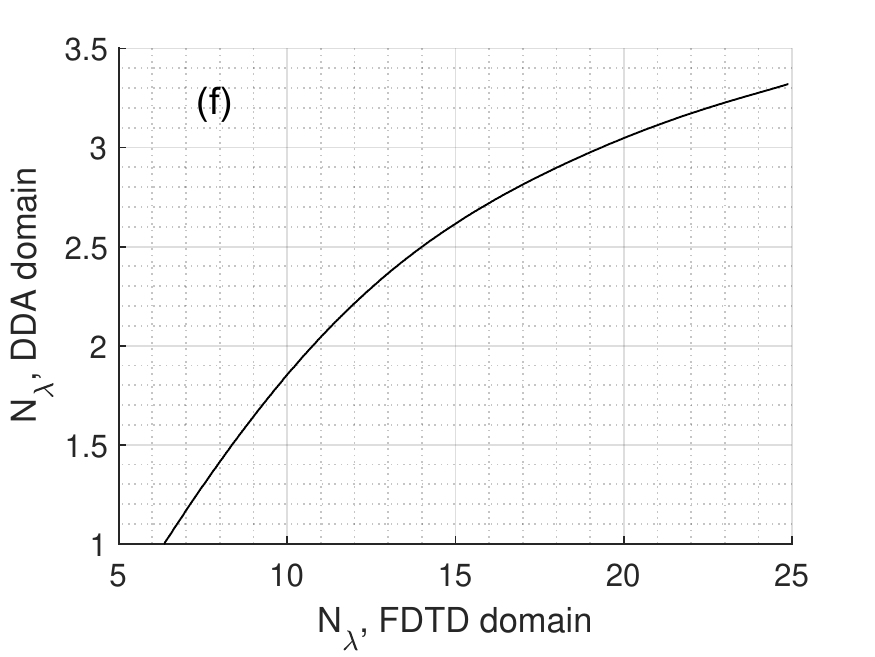}
    \caption{(a) The error of the far-field projection along $\theta\in(0,180)$ given $\phi=0$ in terms of $\mathrm{S}_{11}$. (b) The corresponding compute time in core-seconds, and (c) the total storage memory required for the solution, in Giga-bytes (GB). The comparison is made assuming that separate core DDA simulations are independent and operate as a single process only. The fitted functions are compared numerically such that ratios of (d) compute time and (e) memory are estimated at the same accuracy for a single core process. These ratios are consistent with (f) the number of grid units required for similar accuracy solution in each method.}
    \label{fig:8}
\end{figure*}
We explore different dependencies of the near-field errors (Fig. \ref{fig:6}) and far-field errors (Fig. \ref{fig:7}) on the size of the grid with respect to the same size sphere of radius $r=4.0095$ \si{\micro m}. In each case, Kevin G. Zhu's\cite{Zhu_2020} implementation of the Mie solution is used to evaluate an analytical $\mathrm{S}_{11}$ throughout the volume of the near field, whereas Philip B. Laven's MiePlot v4.6.21 is used to represent $\mathrm{S}_{11}$ along a single plane of incidence in the far field \cite{laven2003simulation}. Refractive index of the sphere and its background are studied for a single case of $n=1.39+i0.01$ with background refractive index $n_b=1.33$. In order to examine the effect of grid size on the scattering pattern, the number of units per wavelength is increased from $N_\lambda=1.49$ to $N_\lambda=9$ in DDSCAT 7.3.3, and from $N_\lambda=10$ to $N_\lambda=25$ in ANSYS\textsuperscript{\textregistered} Lumerical FDTD. 

DDSCAT 7.3.3 remains numerically stable at $N_\lambda\approx1.5$, as a DDA finite element method, but requires the small loss coefficient ($\kappa=0.01$) in order for the CCG solution to converge. Fig. \ref{fig:6} shows that although $\mathrm{S}_{11}$ does not adequately represent analytical Mie theory in the near field when $N_\lambda\approx1.5$, the mean absolute numerical error decreases as $N_\lambda$ increases. However, the relation is not a simple function. Lattice dispersion and geometric approximations affect the numerical accuracy separately. Lattice dispersion occurs as a consequence of representing point dipoles with a  finite separation as opposed to an analytically infinitesimal separation from continuum electrodynamics. Geometric approximation error occurs as an under-representation of a boundary separating two types of point dipoles. Each type of error is distinguished by its effect on the analytical Green's function, where the lattice dispersion is corrected in the DDA \cite{draine1994discrete}, but the geometric approximation error is not. Although the rate of growth in computational time and memory with $N_\lambda$ exceeds ANSYS\textsuperscript{\textregistered} Lumerical FDTD, Fig. \ref{fig:8}(a) shows that the FFP of $\mathrm{S}_{11}$ rapidly approaches the analytical Mie solution within $N_\lambda<5$ such that it requires an order of magnitude less time and memory than ANSYS\textsuperscript{\textregistered} Lumerical FDTD to achieve a solution for $\mathrm{S}_{11}$ of similar accuracy.

In ANSYS\textsuperscript{\textregistered} Lumerical FDTD, the accuracy of $\mathrm{S}_{11}$ in the FFP increases with the grid resolution, as per the Courant Friedrichs Lewy (CFL) criterion \cite{taflove2005computational} and the corresponding numerical dispersion relation. At only $N_\lambda=10$, the deviation between the FFP predicted by the YL-FDTD method and analytical Mie theory is proportional to the refractive index of the background medium $n_b$ such that $\theta_{\mathrm{lum}} \approx \theta_{\mathrm{true}}/n_b$. Fig. \ref{fig:7}(a) shows that the FFP of $\mathrm{S}_{11}$ approaches $S^{\textrm{Mie}}_{11}$ as $N_\lambda$ is increased, but maintains an angular offset regardless of $N_\lambda$. The offset increases with the scattering angle from the direction of propagation to 180$^\circ$, although the general pattern is retained regardless of $N_\lambda$. In contrast, Fig. \ref{fig:7}(b) shows that the DDA preserves the angular representation of the FFP when compared with Mie theory, yet the magnitude of $\mathrm{S}_{11}$ deviates due to the increased near-field error on the smaller grid. Fig. \ref{fig:8} shows that, ultimately, the angular deviation of $\mathrm{S}_{11}$ contributes greater error to the YL-FDTD method FFP resulting in increased time and memory requirements to achieve a similar accuracy solution. 

FDTD simulations were repeated for: (1) lossless spheres embedded in dielectric media; (2) lossy dielectric spheres in free space; (3) lossless dielectric spheres in free space; and (4) different automatic meshing and conformal mapping variants for the same scattering sphere geometry and source condition. In each case, the far-field projection demonstrates the same angular offset in disagreement with Mie theory, which suggests that the angular difference is caused from numerical dispersion error alone (refer to Supplementary information). 
\begin{figure*}
    \centering
    \includegraphics[width=0.45\textwidth]{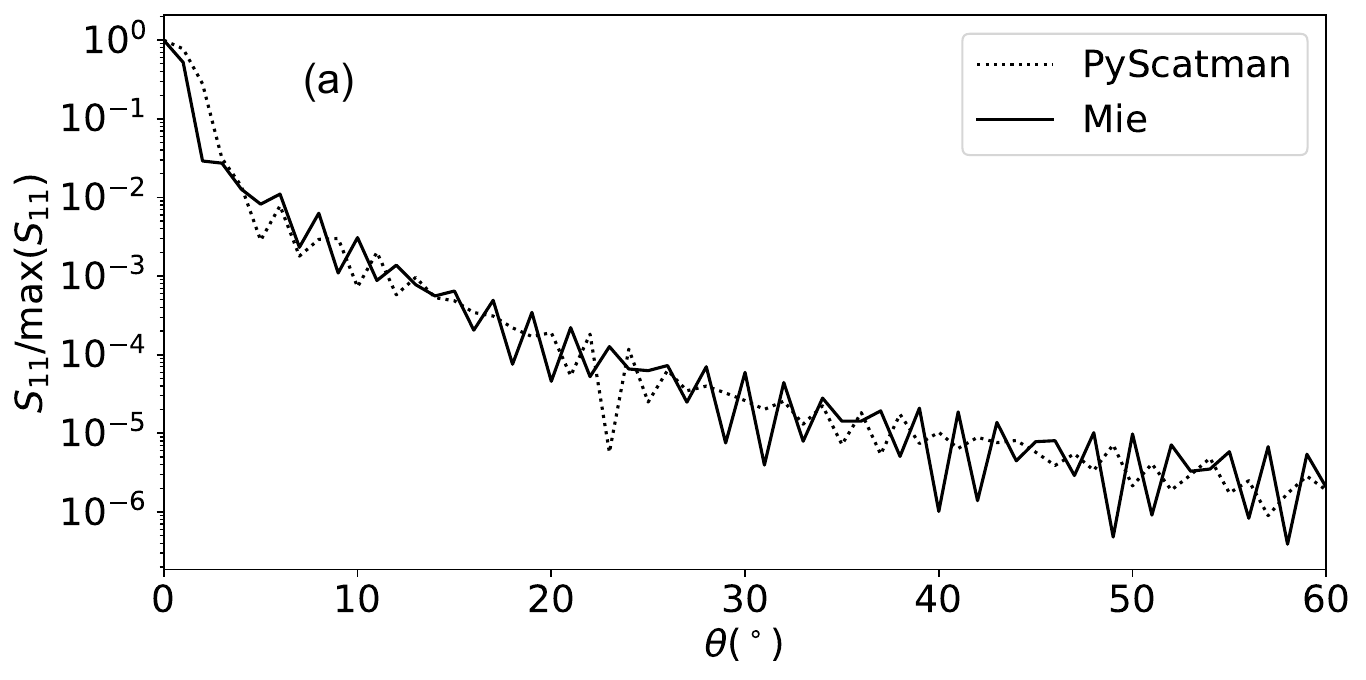}
    \includegraphics[width=0.45\textwidth]{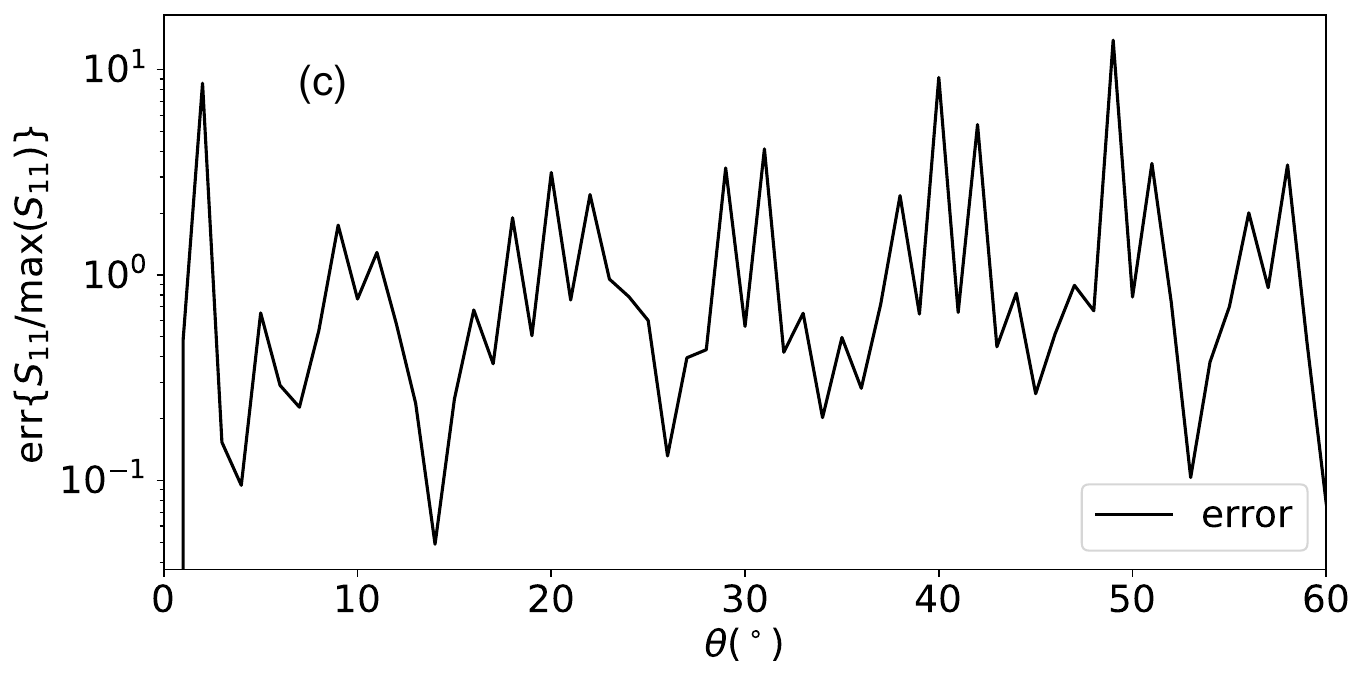} \\
    \includegraphics[width=0.45\textwidth]{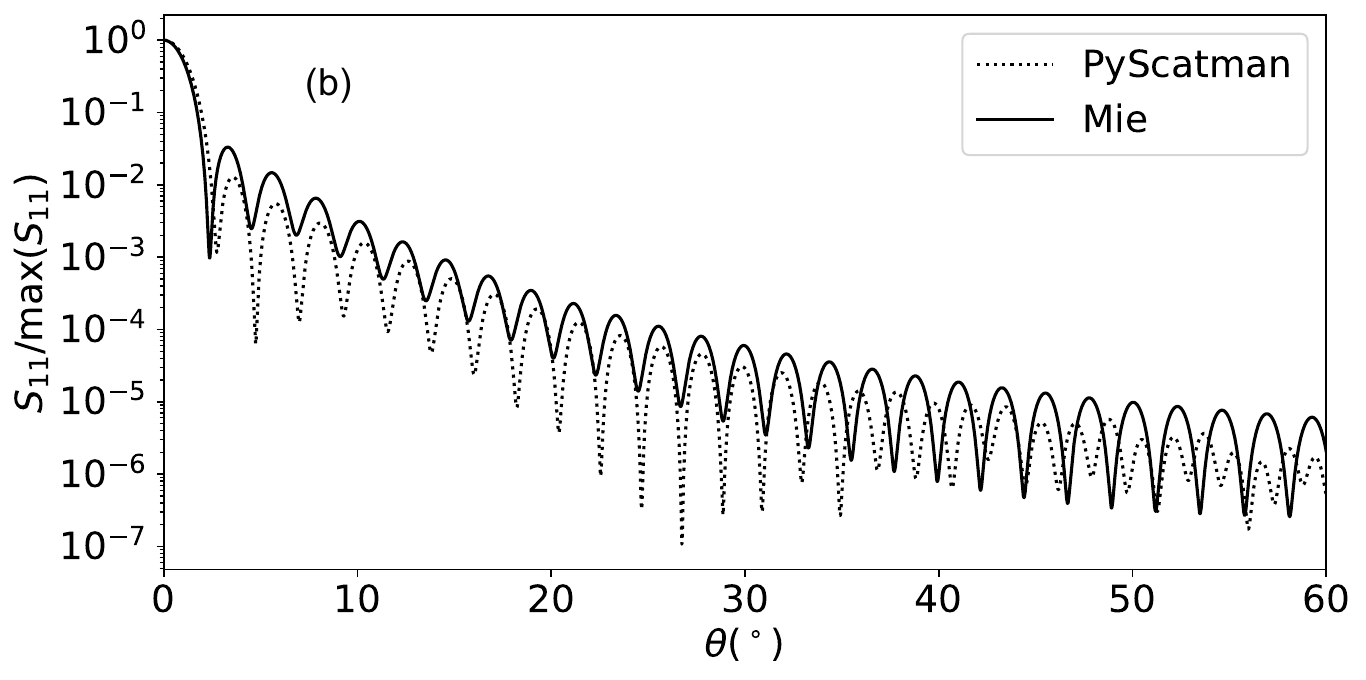}
    \includegraphics[width=0.45\textwidth]{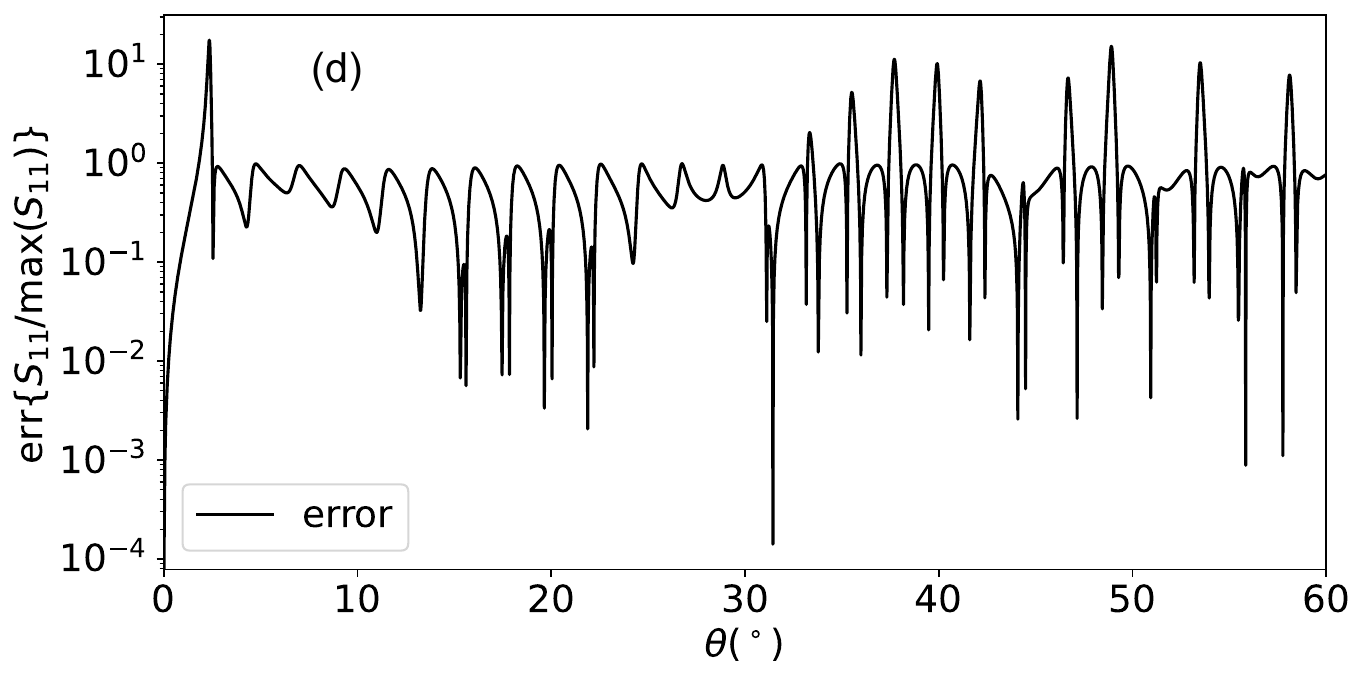}
    \caption{Comparison between MiePlot v4.6.21 and the MSFT method \texttt{PyScatman} given an ideal (angular) detector of the far-field projection: (a) 128$\times$128 detector array; (b) a 4096$\times$4096 detector array. The corresponding absolute error of each solution over the maximum angular range $\theta\in[0,60]$. The error does not change considerably with the size of the grid, which suggests that the error is the same regardless and occurs from the approximation itself. However, the solutions are in qualitative agreement which may be sufficient in the context of pattern recognition systems. }
    \label{fig:9}
\end{figure*}

\begin{figure*}
    \centering
    \includegraphics[width=0.45\textwidth]{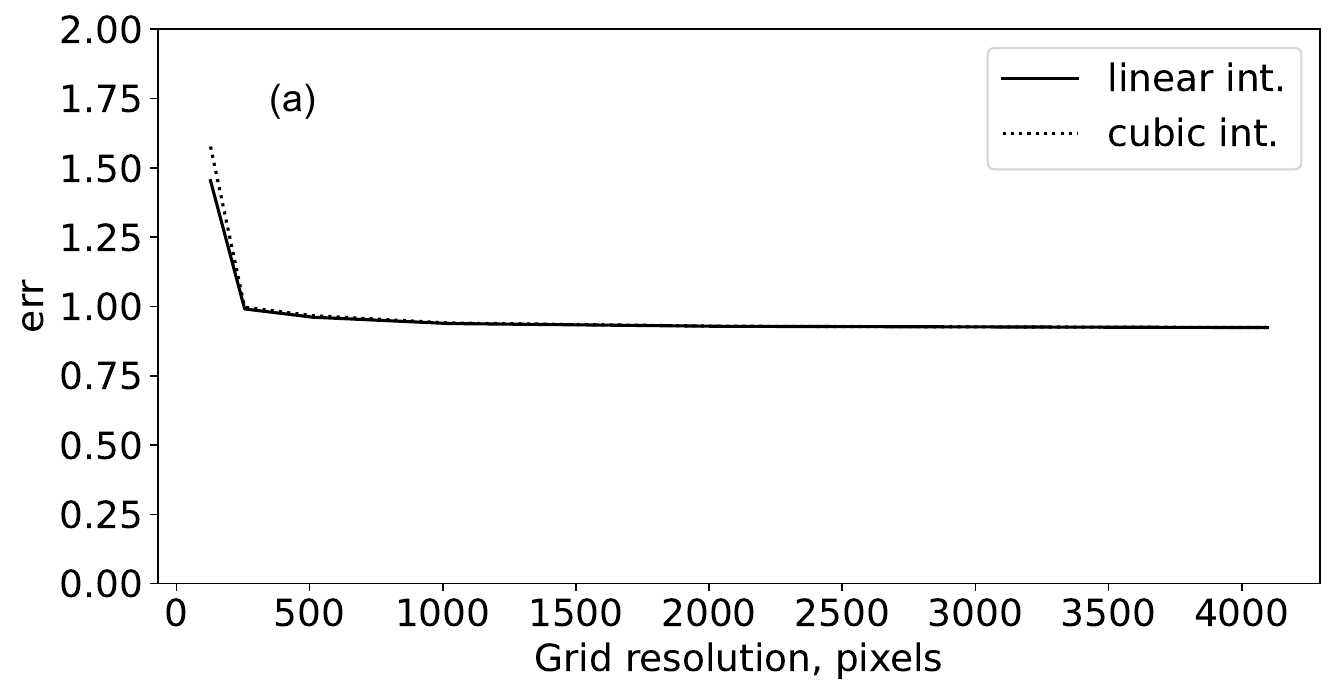}
    \includegraphics[width=0.45\textwidth]{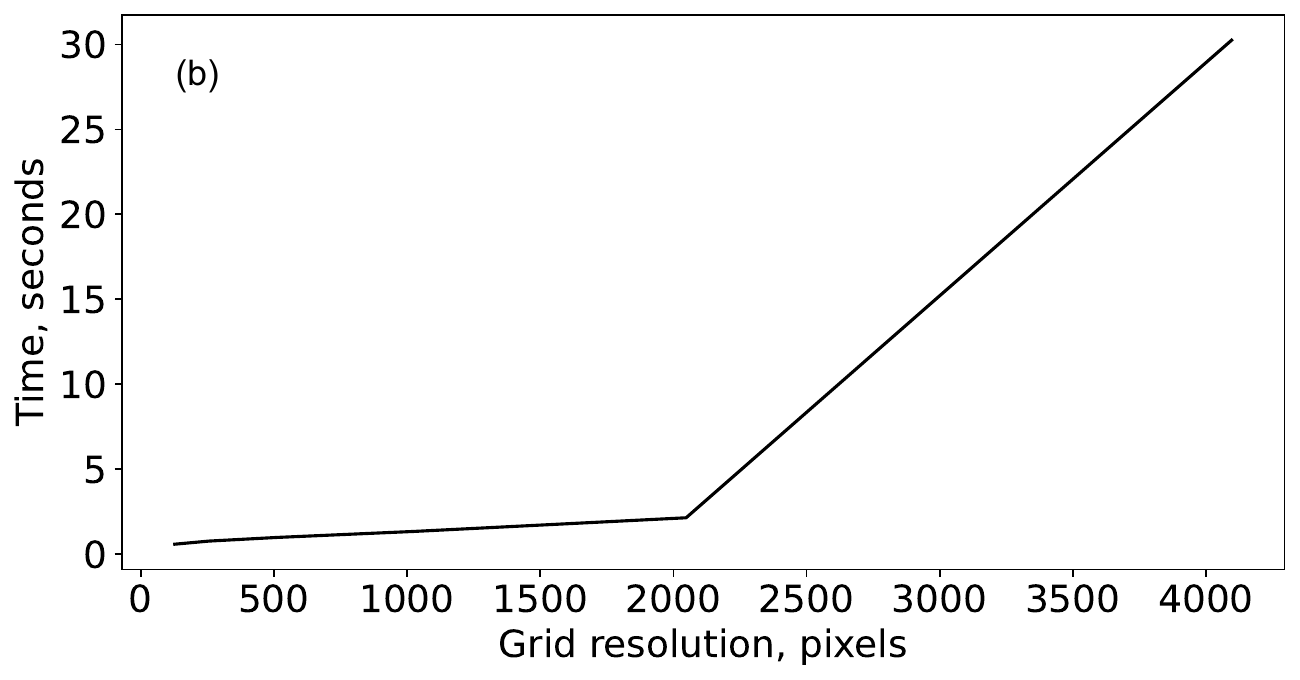} \\
    \includegraphics[width=0.45\textwidth]{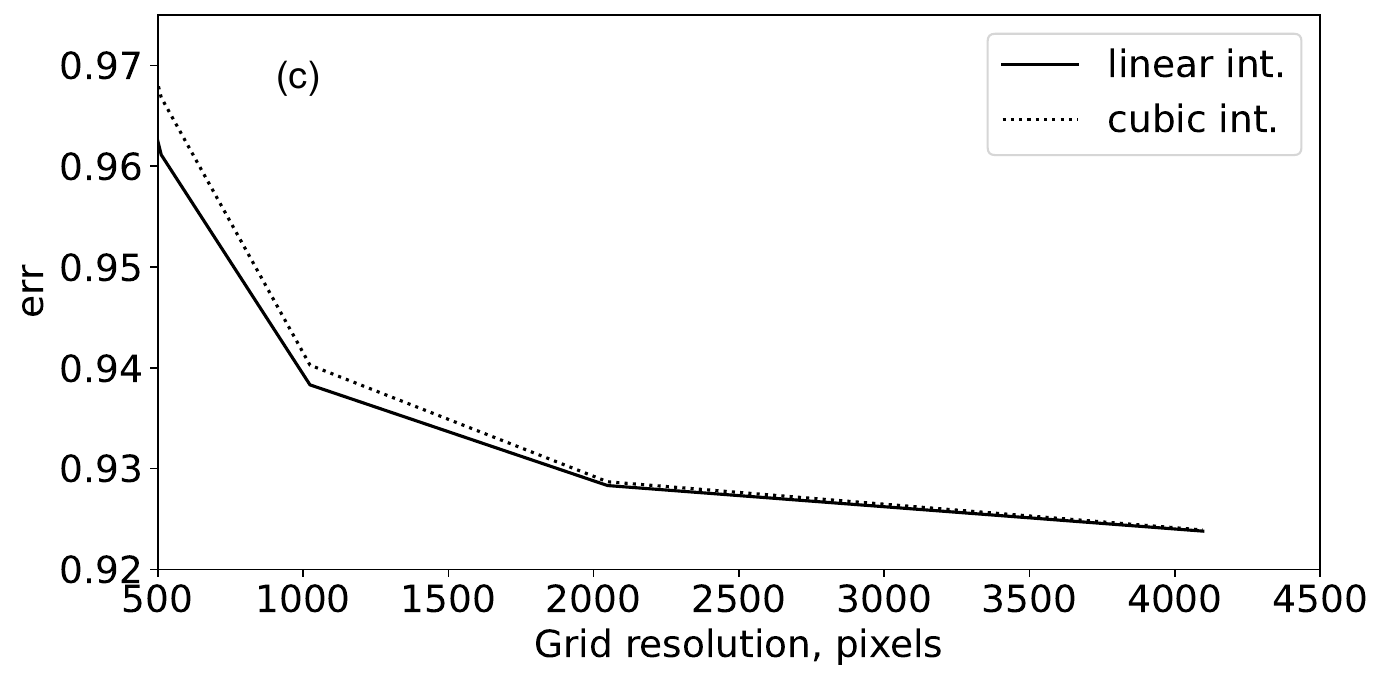}
    \includegraphics[width=0.45\textwidth]{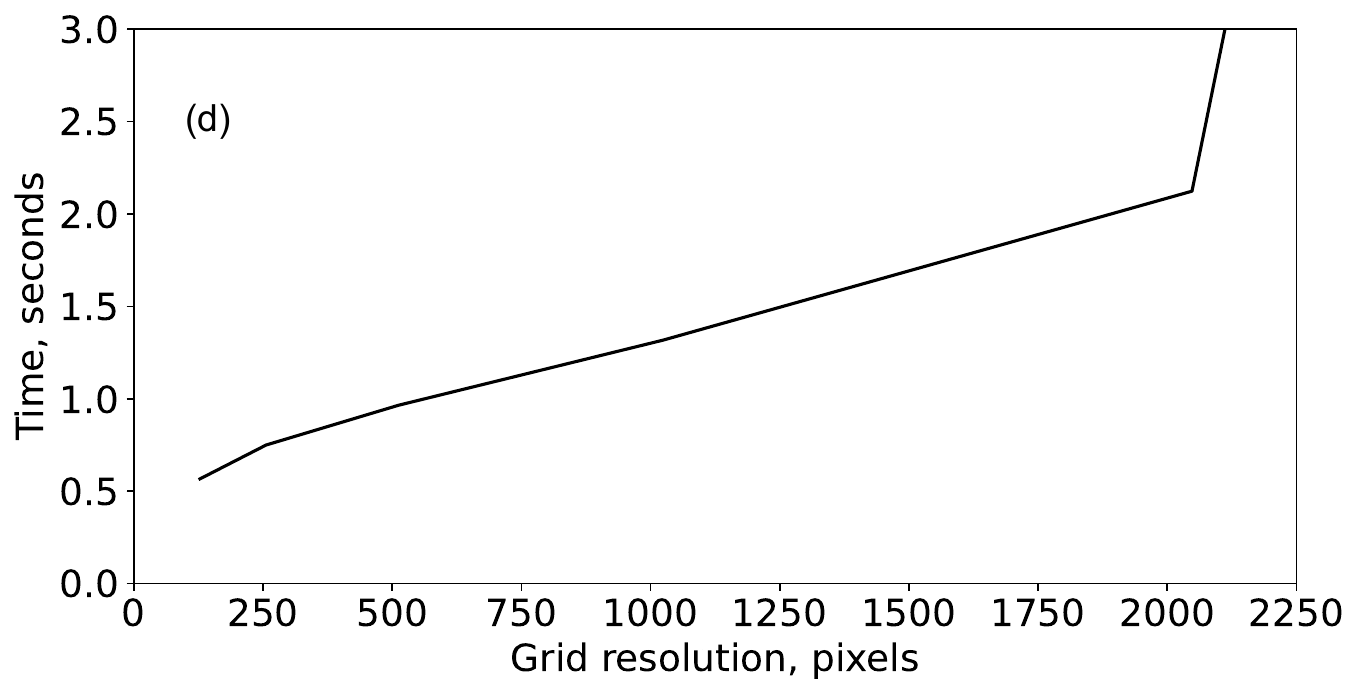}
    \caption{(a) The mean absolute error of the multi-slice Fourier transform Ideal detector projection from \texttt{Pyscatman} compared with MiePlot v4.6.21. (b) The time required for computation on each grid size. The grid is square and the unit along the x-axis is the number of pixels. CUDA was used on an NVIDIA RTX A4500 GPU for performing the fast-Fourier transform. (c) and (d) illustrate the same plots in the domain with least slope. Computational speed and accuracy gains were negligible as the number of pixels was increased, suggesting an asymptotic limit on the achievable accuracy. Therefore, a greater number of simulation cases with increasing spatial resolution approaching the limit of virtual memory was considered redundant.}
    \label{fig:10}
\end{figure*}
Fig. \ref{fig:7} shows that in both the DDA and YL-FDTD method, the relative error of the scattering pattern increases with angular difference in scattering angle, $\theta$, from the direction of incidence. Angular error of $\mathrm{S}_{11}$ in the the YL-FDTD method FFP is distinct from the amplitude error of $\mathrm{S}_{11}$ in the DDA FFP, the latter representing the correct angle for $\mathrm{S}_{11}$ but with a lower amplitude. Angular errors in the YL-FDTD method are of consequence to the positioning of a CMOS camera or photon detector within a narrow angular range for scattering pattern measurement, whereas the amplitude errors in the DDA are of consequence to selecting the appropriate dynamic range of a detector for a scattering pattern measurement. The cause of errors in $\mathrm{S}_{11}$ as $\theta$ increases is rarely emphasized in the literature concerning the measurement of scattering patterns from biological cells, yet remains critical for accurate representation of back-scattering patterns and their analysis. Increases in error with scattering angle are caused by approximations to the far-field projection based on $E_\theta$ and $E_\phi$ only, whereas $E_r$ also contributes to the FFP beyond 90$^\circ$. In fact, $E_r$ is most responsible for wide-angle scattering since it causes the field to attenuate or refract, and re-direct electromagnetic power. This is why the paraxial BRA MSFT method cannot be used to represent electromagnetic scattering for $\theta>60^\circ$. 


Figs. 9-10 show that the \texttt{PyScatman} BRA MSFT method exhibits significant error compared with either the YL-FDTD method or DDA for predicting the FFP of electromagnetic scattering in the angular range  $\theta=[0,60)^\circ$ from a sphere of radius $a_{\mathrm{eff}}=4.0095$ \si{\micro m} and refractive index of $n=1.39+i0.01$ immersed in a background refractive index of $n_b=1.33$. Although the results agree with Mie theory, the error increases with angular resolution, exhibiting a broad range between $10^{-4}$ and 10. In reconstructing a nanometer refractive index, this would seem inappropriate. However, for qualitative comparison with forward elastic light scattering experiments from complex geometries or pattern recognition, it may be significantly faster and sufficient for reconstructing scattering patterns. Such patterns may include interference from microscope lenses or other optics used to measure single-cell light scattering in neutral pH index-matched environments, or flow cytometry, where the paraxial approximation and beam-envelopes are commonly used. However, the inaccuracy of the method makes it inappropriate as a replacement for full-field simulation for broad angle scattering. Additional limitations to the approach include its inability to represent near-field scattering behaviour, which is accomplished directly in both the YL-FDTD method and DDA method.

\subsection{Far-field scattering from S. \emph{cerevisiae}}

\begin{figure}
    \centering
    \includegraphics[width=0.45\linewidth]{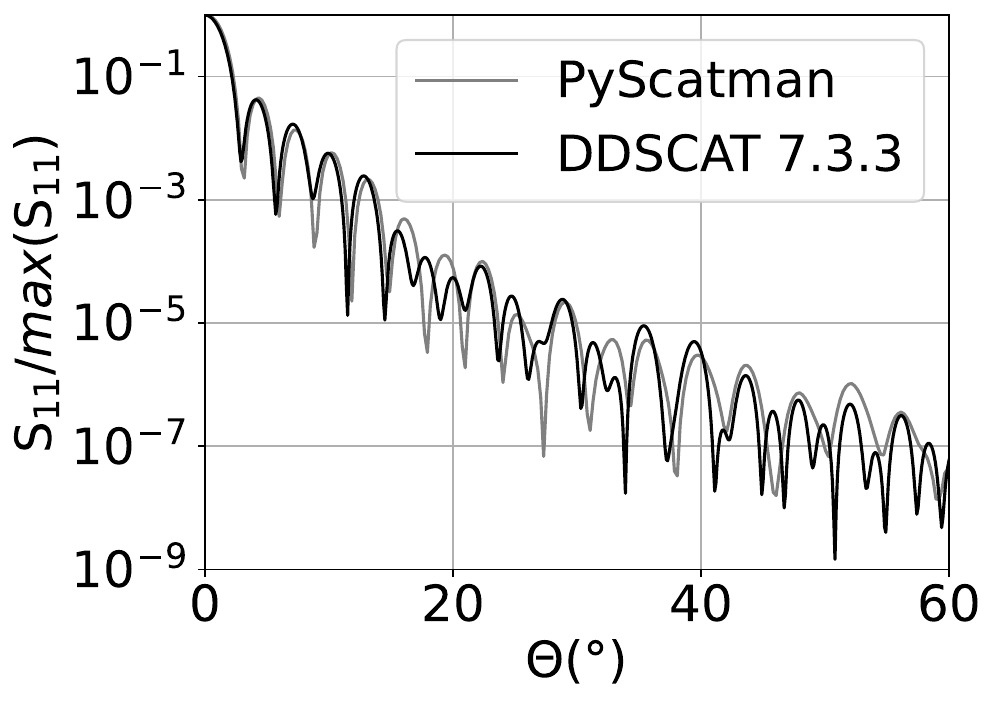}
    \includegraphics[width=0.45\linewidth]{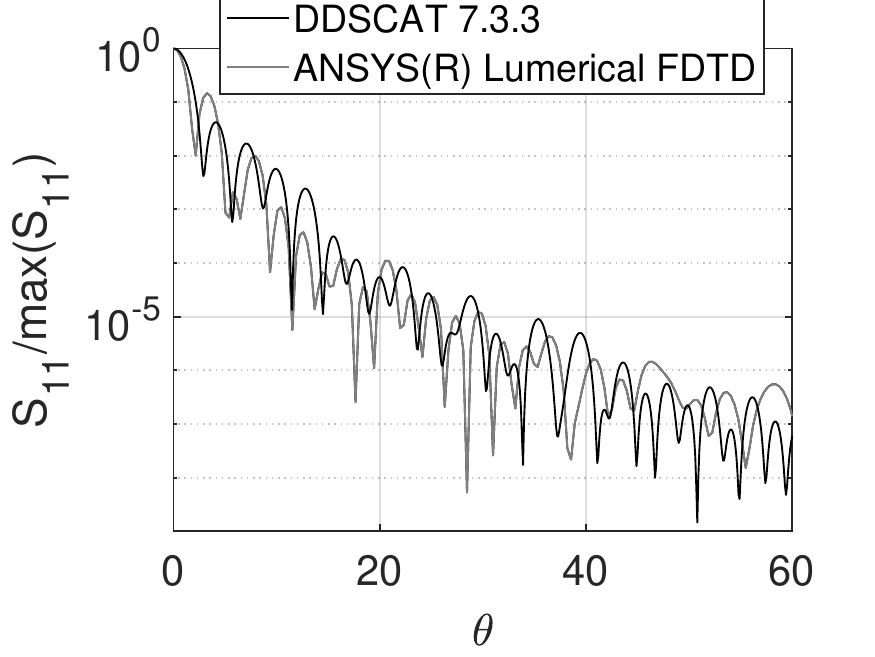}
    \caption{(a) PyScatman plotted with DDSCAT 7.3.3 $S_{11}$ in the angular range $\theta\in[0^\circ,60^\circ)$. (b) DDSCAT 7.3.3 plotted with ANSYS\textsuperscript{\textregistered} Lumerical FDTD in the angular range $\theta\in[0^\circ,60^\circ)$. There is better agreement between the plots in (a) than in (b), suggesting that the Born-Rytov approximation is more accurate in representing electromagnetic scattering from the complex geometry of S. \emph{cerevisiae} produced by tomographic phase microscopy \cite{habaza2015tomographic}.}
    \label{fig:11}
\end{figure}

Fig. \ref{fig:11} shows a similar comparison of $S_{11}$ scattered along a single plane ($\theta=[0^{\circ},60^{\circ})$, $\phi=0^{\circ}$) from $\lambda_0=405$ \si{nm} light incident on S. \emph{cerevisiae}. Given the loss of momentum along sharp boundaries in the Yee-lattice algorithm used in ANSYS\textsuperscript{\textregistered} Lumerical FDTD, the near-field to far-field projection predicts a different intensity pattern ($S_{11}$) from DDSCAT 7.3.3 along this plane. In contrast, the \texttt{PyScatman} intensity pattern for $S_{11}$ is similar to the DDA's, which was previously shown more accurate at representing angular scattering than the YL-FDTDM (Fig 8). This result demonstrates how a paraxial approximation can be better than commercial FDTD for representing forward electromagnetic scattering from heterogeneous media representative of measured biological cells, particularly if a cell exhibits significant spatial variation or sharp boundaries in its refractive index along the direction of propagation.

\begin{figure*}
    \centering
    \includegraphics[width=0.3\textwidth]{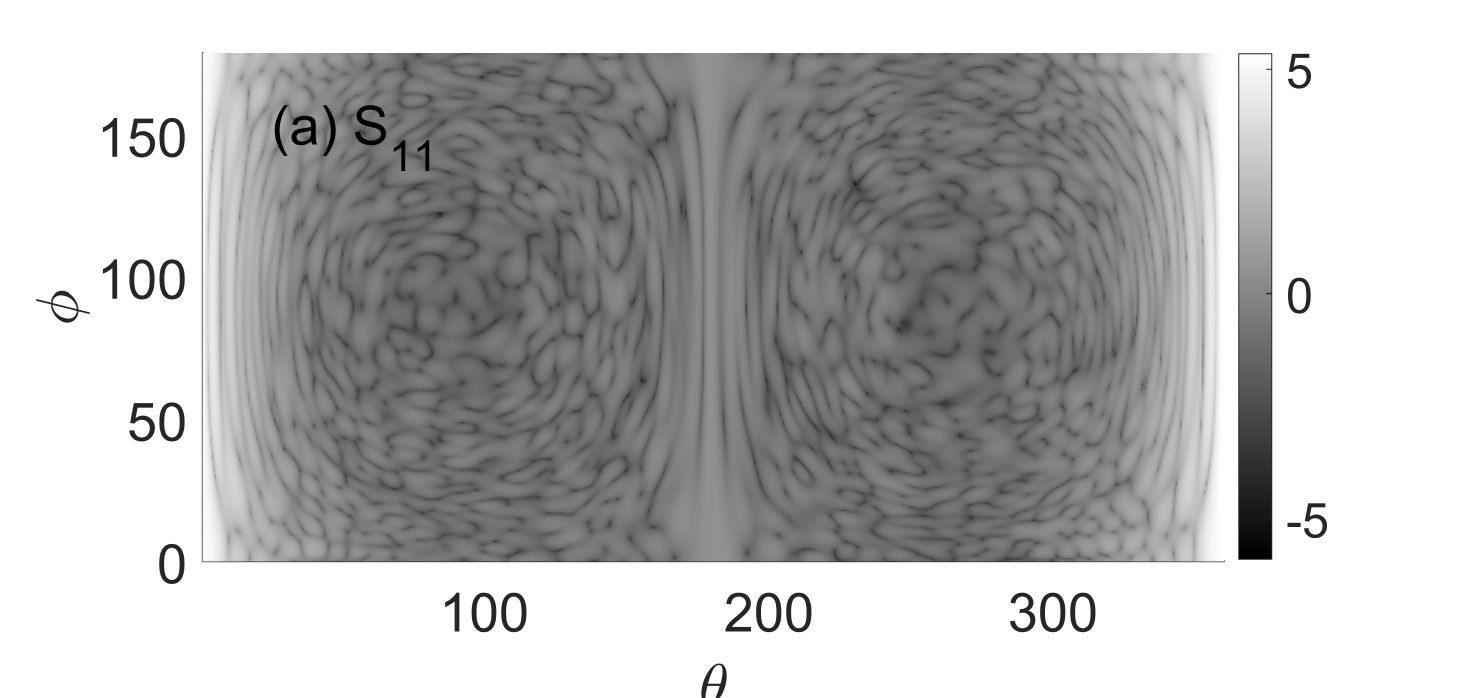}
    \includegraphics[width=0.3\textwidth]{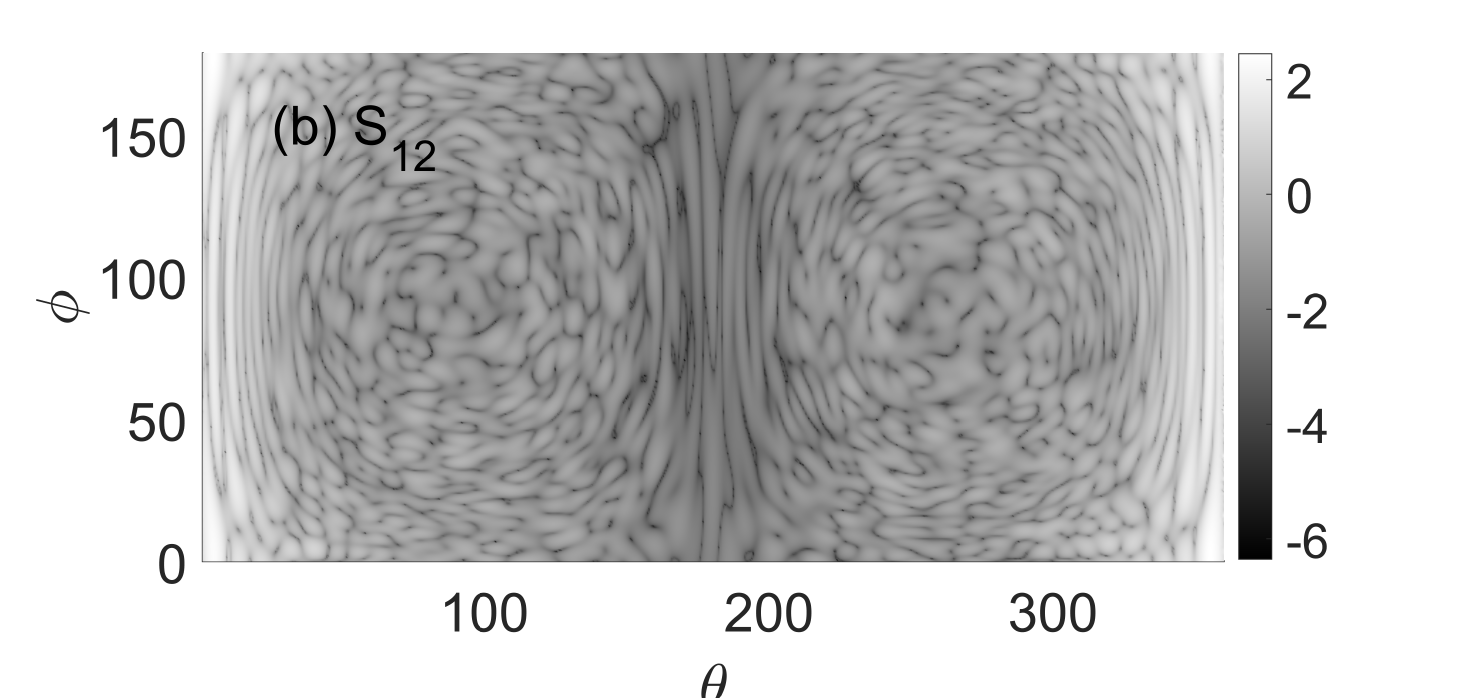}
    \includegraphics[width=0.3\textwidth]{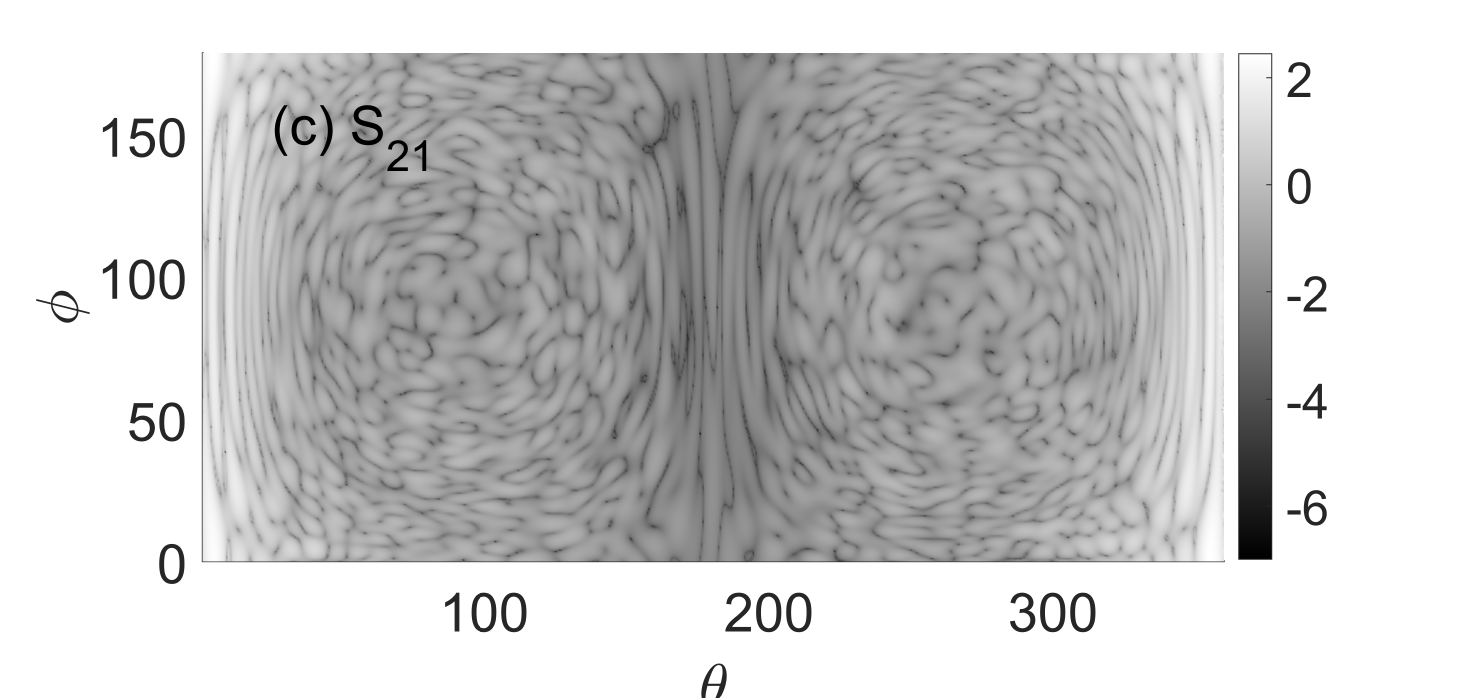} \\
    \includegraphics[width=0.3\textwidth]{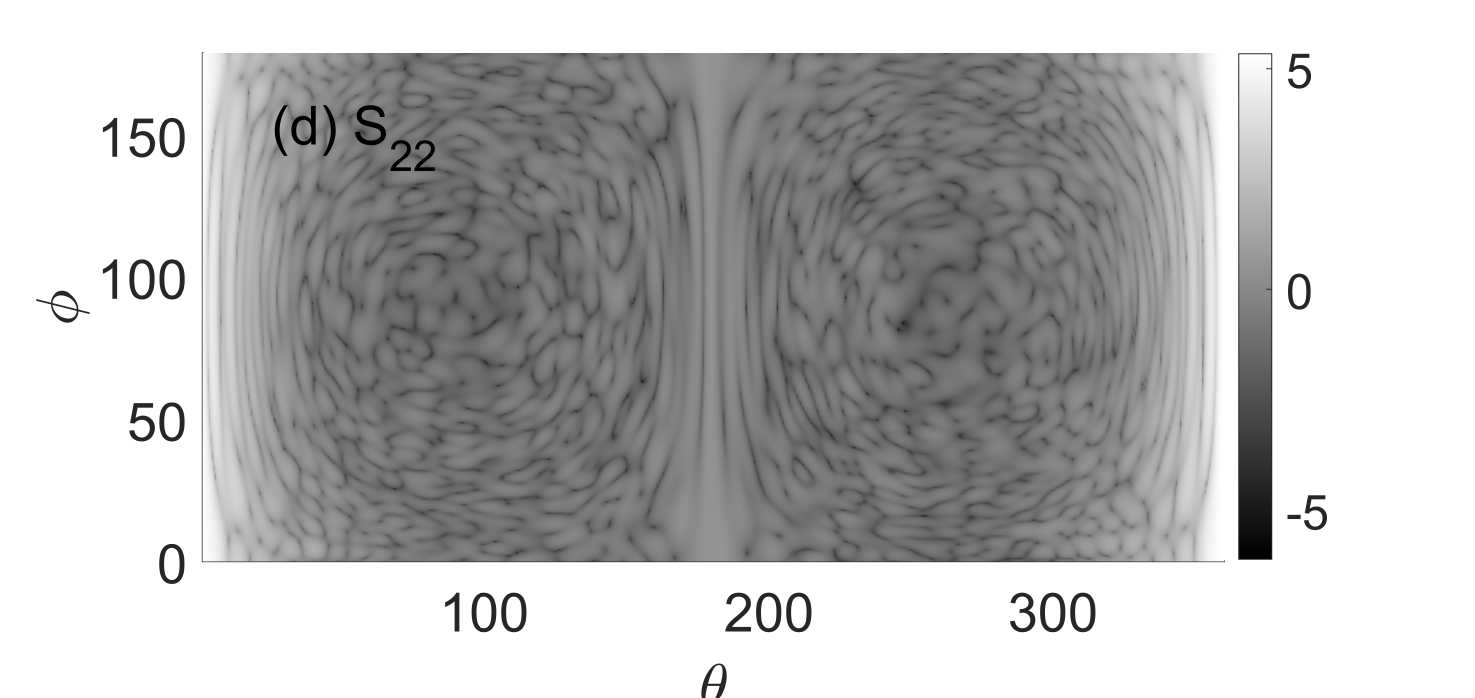}
    \includegraphics[width=0.3\textwidth]{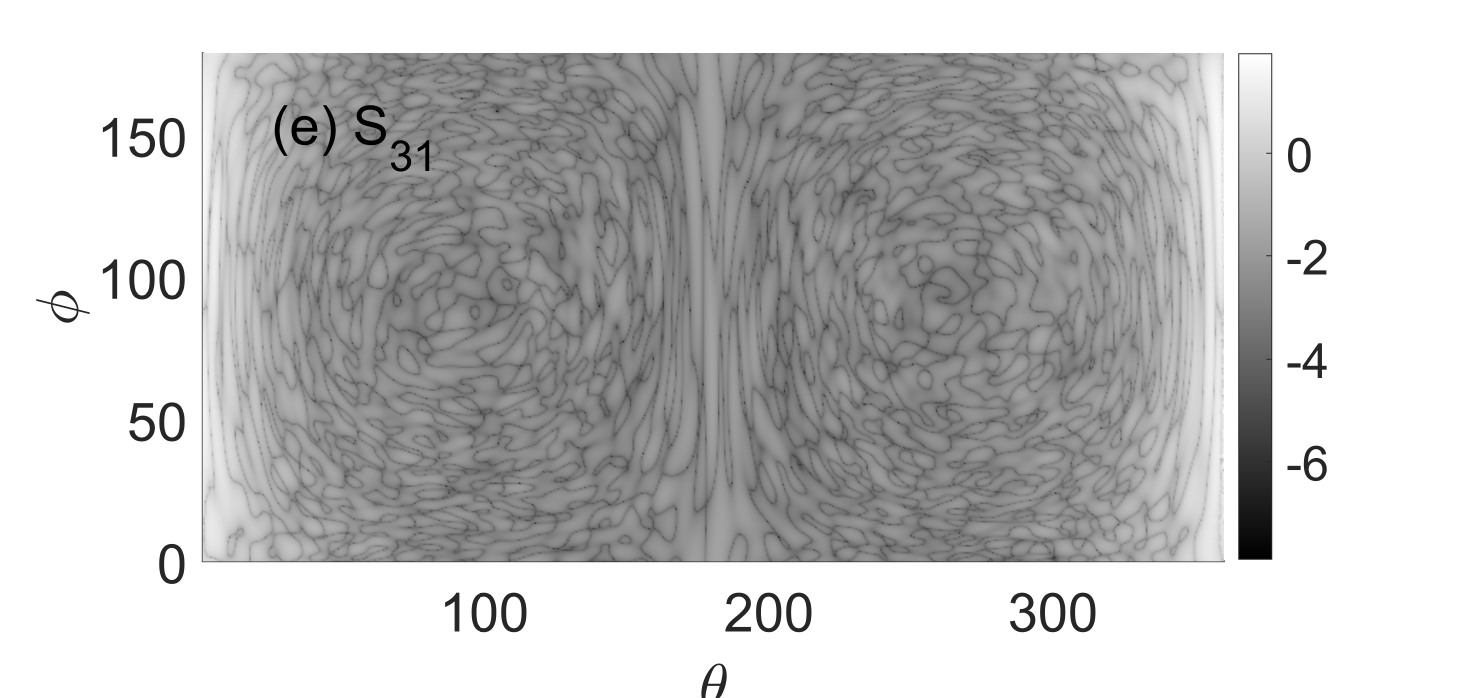}
    \includegraphics[width=0.3\textwidth]{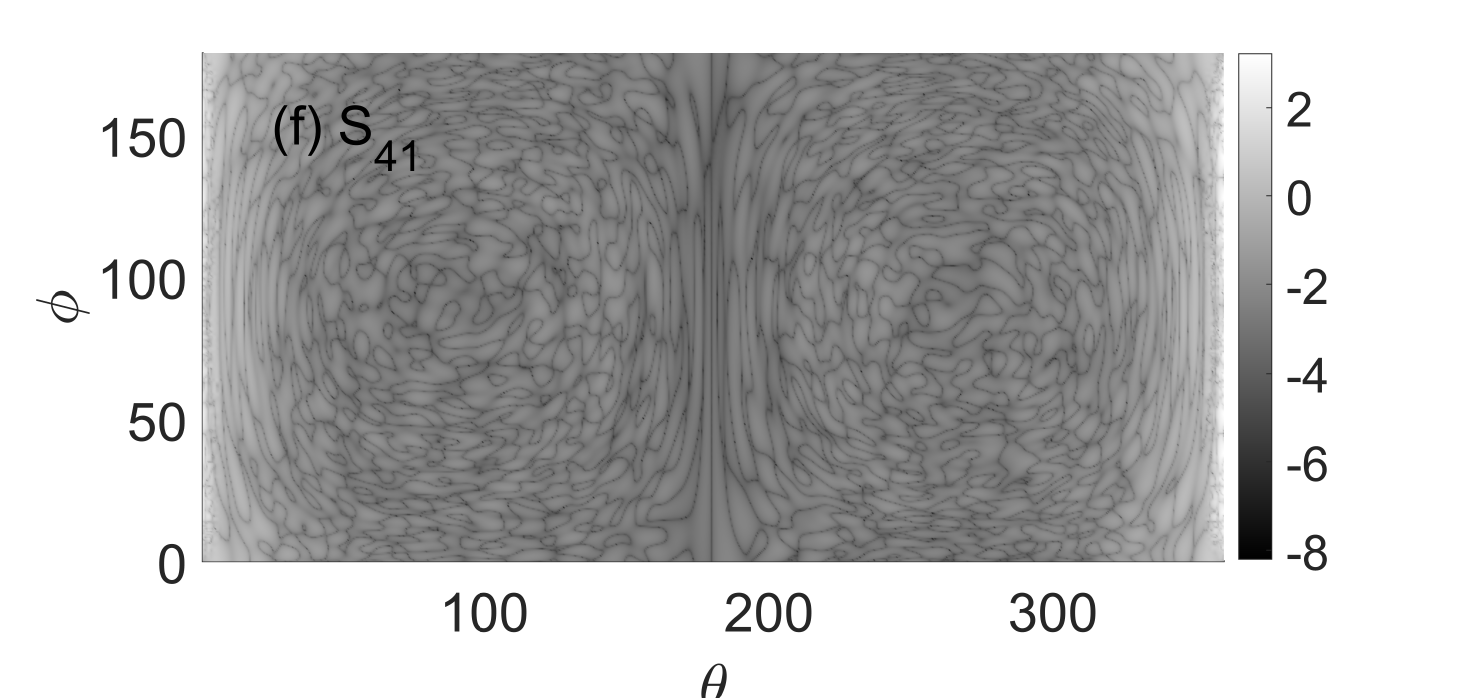}
    \caption{Scattering coefficients plotted for each cell in logarithmic units, each normalized with respect to peak intensity. Notice that the scattering solution is asymmetric, such that the complete spherical scattering pattern is required to fully reconstruct the geometry and optical properties from the field intensity alone.}
    \label{fig:12}
\end{figure*}

\begin{figure*}
    \centering
    \includegraphics[width=0.75\linewidth]{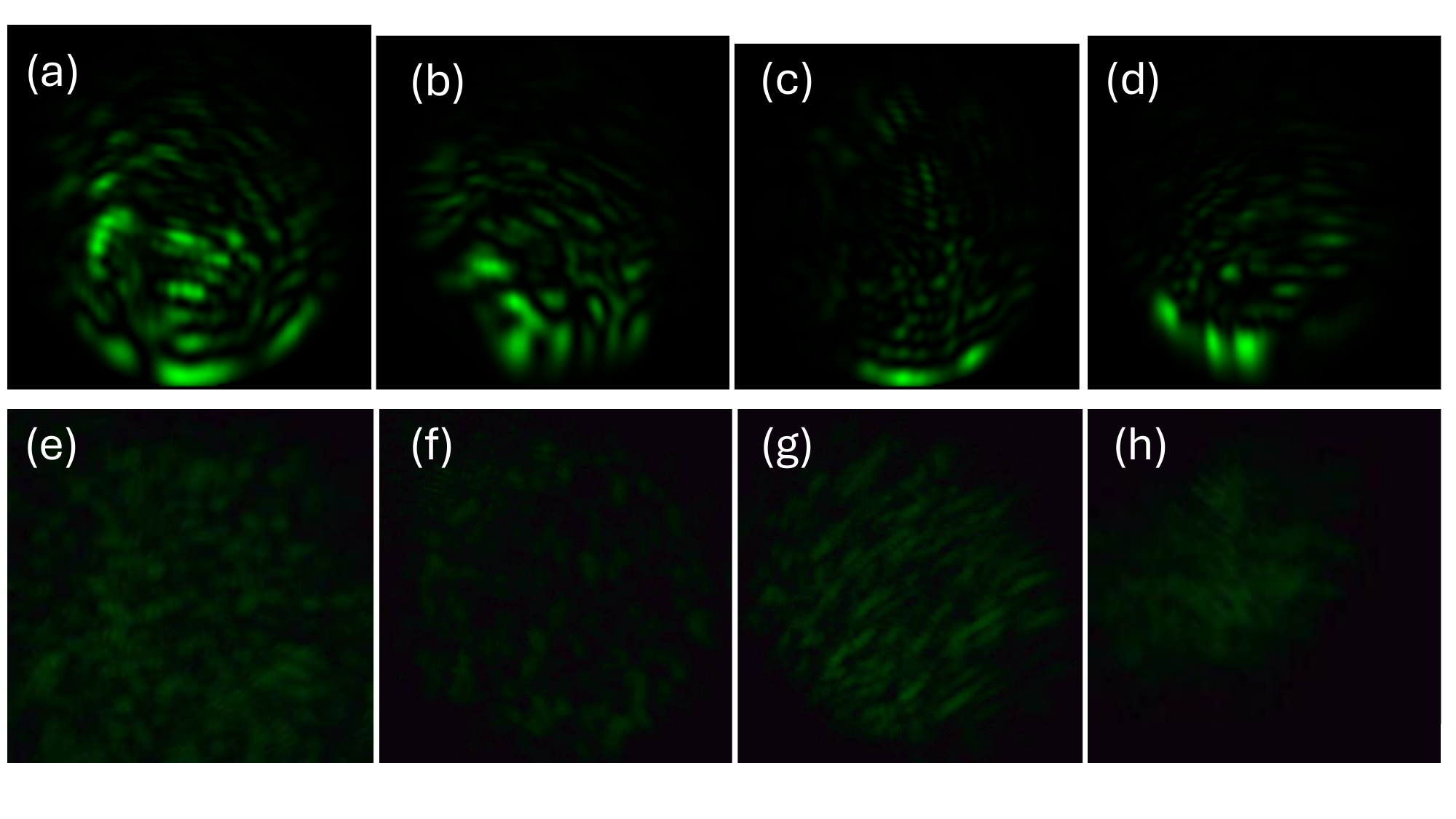}
    \caption{(a-d) Simulated electromagnetic side-scattering patterns from the same S. \emph{cerevisiae} reconstruction, but with an incident wavelength of $\lambda_0=532$ \si{nm} (similar to Habaza \emph{et al.}'s \cite{habaza2015tomographic}) and variation in orientation with respect to an incident laser far-field air-water interface. (e-h) Empirically measured side-scatter patterns from a yeast cell after 24 hours, using a $\lambda=532$ \si{nm} laser diode. Dissimilarities in the patterns occur since the scattering does not occur from the exact same yeast cell, yet the patterns have a likeness that can be used for machine-learning classification  purposes \cite{liu2023multi}. }
    \label{fig:13}
\end{figure*}

Fig. \ref{fig:12} shows that complete $4\pi$ \si{str} far-field scattering pattern from a biological cell can be obtained for polarization and up to twelve Mie scattering coefficients in DDSCAT 7.3.3. In contrast, ANSYS\textsuperscript{\textregistered} Lumerical FDTD may only return the far-field scattered light intensity, p-polarized and s-polarized electric field components $E_p$ and $E_s$ and/or their phase, which must be related to the scattering coefficients. Each near-field to far-field transform can be implemented with an interface, which illustrates the critical angle cut-off of the patterns as they are measured from the far-field (Fig. \ref{fig:13}). Each numerical method provides a tool for modeling electromagnetic scattering from biological cells, but with different angle-dependent accuracy and performance which is critical for gauging the success and maximum accuracy for machine-learning classification \cite{liu2023multi}. Comparing each numerical method's far-field projection can allow for both qualitative and quantitative assessment of Born-Rytov approximations, which assists in testing both improved three-dimensional scattering solutions for Maxwell's equations, or effective refractive index reconstruction of biological cells such as S. \emph{cerevisiae}.  This approach may also provides a means of assessing the accuracy of paraxial electromagnetic scattering from heterogeneous refractive index, which is important for modeling background media and the signal quality of optical systems. Analytical Mie scattering theory may be an insufficient test for these problems, since electromagnetic scattering from a sphere is largely symmetric and fails to represent electromagnetic scattering asymmetries that can occur from sharp corners or boundaries of reconstructed cell geometries \cite{ warner2023comparative}. Simulations of these type are also useful for label-free cytometry instrumentation, where the angular distribution of scattered light can be used to assess the detector angle and distance from the scattering region to satisfy required signal to noise ratio and dynamic range of the detector. 

Fig. \ref{fig:12} illustrates the 4$\pi$ scattering pattern obtained from an effective refractive index reconstruction by Habaza \textit{et al.}\cite{habaza2017rapid}. In the center of each plot is the azimuthal scattering plane in which the scattered intensity is weakest. On either side of this boundary there is an asymmetry in the pattern since the geometry and refractive index of the S. \textit{cerevisiae} cell is asymmetric. Consequently, a single angular range that is less than $4\pi$ \si{str} provides only partial information or reconstruction of the cell. This is why Habaza \textit{et al.}\cite{habaza2017rapid} employed an optical trap to provide projections from multiple angles for the reconstruction. Six different scattering coefficients were obtained using DDSCAT 7.3.3., yet up to twelve can be requested. Scattering coefficients provide unique features associated with the polarization of light scattered from biological cells, which when measured at different angles has obtained higher accuracy real-time classification accuracy for microalgae \cite{zhuo2022machine}. Similarly, different refractive index reconstructions have been measured from either polarization or phase imaging within the same optical system \cite{wang2024imaging}, which suggests that each measurement only provides partial information regarding the spatial distribution of refractive index. Each study suggests that both electromagnetic wave polarization and phase are significant for refractive index reconstruction, and electromagnetic scattering pattern measurements for cell classification.

\section{Conclusion}
In this work we have compared the far-field scattering pattern from the finite-difference time-domain (FDTD) method, the discrete dipole approximation (DDA) method, and the Born-Rytov approximation multi-slice Fourier transform (BRA MSFT) method to the case of an analytical sphere and a geometrically complicated tomographic reconstruction of a biological cell, S. \emph{cerevisiae}, with heterogeneous refractive index. Although each numerical method is comparable with analytical Mie theory, only the DDA and BRA MSFT are in agreement for modeling forward scattering of visible wavelengths from S. \emph{cerevisiae}. Of the three methods, both DDA and FDTD is applicable for wide-angle light scattering from S. \emph{cerevisiae}, but only DDA provides appreciable quantitative accuracy in contrast to the case of spherical scattering. Comparisons for a sphere demonstrate the theoretical error of the Born-Rytov approximation applicable to the tomographic reconstruction. After comparison with analytical Mie theory, the DDA is best suited for accurate assessment of the Born-Rytov approximation error for far-field scattering from biological cells, potentially given momentum continuity errors at sharp boundaries in FDTD.

Although the BRA MSFT method introduces numerical error, it shares better qualitative agreement with the DDA than FDTD within the forward scattering cone angle of $\theta\in[0,60)^\circ$, and its rapid evaluation of the electromagnetic scattering pattern is useful for assessing the dynamic range and separation distance of photon detectors in forward scattering flow cytometry. Despite operating orders of magnitude faster than DDA, DDSCAT 7.3.3 can achieve similar accuracy 90$^\circ$ far-field scattering patterns with an order of magnitude less error than the YL-FDTD method far-field projection, given the latter's numerical dispersion error. Accurate far-field electromagnetic scattering patterns predicted by the DDA are useful for cell reconstruction and machine learning classification of biological cells. However, the DDSCAT 7.3.3 implementation allows only a small variation in refractive index throughout the spatial domain and background, whereas the YL-FDTD method has the particular advantage of representing greater differences in effective refractive index and a near-field time domain solution with non-uniform spatial resolution. However, significant variation in refractive index in the visible range is not typical of biological cells. Typical of most finite element solvers, the DDSCAT 7.3.3 DDA also requires a small loss to be added to the cell in order to converge, whereas background loss in the YL-FDTD method may not be fully represented in the far-field projection. Another limitation to commercial FDTD solvers, such as ANSYS\textsuperscript{\textregistered} Lumerical FDTD, is the geometric definition of complicated geometries such as S. \emph{cerevisiae} in a stereolithographic (STL) mesh format. This can cause incongruities in the geometry, adding to the numerical dispersion and momentum conservation error at material boundaries, which already limit the accuracy of the far-field projection in the case of spherical scattering. It seems likely that the DDA and BRA MSFT are in better agreement due to not only the YL-FDTD method numerical dispersion, but irregularities in the shape from the STL file format imposed by the commercial solver. Therefore, the DDA and BRA MSFT are more amenable to a forward far-field projection than the YL-FDTD method for the heterogeneous geometry of a biological cell. However, this does not consider more advanced time-domain solvers of Maxwell's equations.  

Although we consider the spatial variation of effective refractive index reconstructions to a sub 6 \si{nm} resolution, as reported by Habaza \emph{et al.}\cite{habaza2015tomographic}, it should be noted that the effective refractive index of the tomographic reconstruction may not provide biological properties to this characteristic length. Effective index neglects cell dynamics, loss, dispersion, and nonlinear scattering. Caution must be employed for the use of tomographic reconstructions of refractive index, from either phase or polarization, to estimate design parameters in flow cytometers using different wavelengths or biological cells where such refractive index reconstructions may vary. However, the DDA would appear an acceptable and useful alternative to FDTD for whole cell reconstruction and far-field simulation of light scattering from biological cells for the purposes of machine learning classification.

\section*{Acknowledgment}
This work was supported by the Natural Sciences and Engineering Research Council of Canada (NSERC) and Alberta Innovates. 
Special thanks to the Ying Y. Tsui lab for use of ANSYS Optics and a side-scatter measurement apparatus.   

\pagebreak
\begin{center}
\textbf{\large Supplemental Materials: Comparing time and frequency domain numerical methods with Born-Rytov approximations for far-field electromagnetic scattering from single biological cells}
\end{center}
\setcounter{equation}{0}
\setcounter{figure}{0}
\setcounter{table}{0}
\setcounter{page}{1}
\makeatletter
\renewcommand{\theequation}{S\arabic{equation}}
\renewcommand{\thefigure}{S\arabic{figure}}

\section{Summary}
This supplementary document reviews: (1) the method for acquiring data from Habaza \emph{et al.} \cite{habaza2015tomographic}; (2) the method of refractive index value truncation and reconstruction; and (3) a comparison of scattering from gradually increasing diameter spheres and different mesh algorithms for the FDTD method.

\section{Automated image processing reconstruction}
MATLAB scripts are provided for the image processing. Habaza \emph{et al.}, provide useful videos of the reconstruction. Over the frames of the video, the complete geometry can be reconstructed using basic image processing. Below is the MATLAB script used to process the images. The section of the image representing the cell, and the colorbar axis, were each manually defined. The full script including re-alignment for the simulation domains is also provided.
\tiny{
\begin{verbatim}
    %% Program for testing the image and video extraction from Habaza et al.,2015
% Ref. for video files:
% This script must be placed in the same directory as the video, and the
% directory named videos_temp/

inputFolder1='videos_temp/';
outputFolder1='images_temp/cell1/'; % Define folder directory of video images
outputFolder2='images_temp/cell1_3d/'; %Define folder directory of figure images
outputFolder3='figures_temp/cell1/';   %Define folder directory of actual figures
outputFolder4='data_temp/';
outputFolders=[string(outputFolder1),string(outputFolder2),string(outputFolder3),string(outputFolder4)];
for d=1:length(outputFolders)
try
    pd=pwd;
    cd(outputFolders(d));
    cd(pd);
catch
    mkdir(outputFolders(d));
end
end

v1=VideoReader([inputFolder1 'ol4081881m001.avi']); % Define video object
vid1Frames=read(v1);        % Read list of frames from video
for f=1:size(vid1Frames,4)  % for each frame in vid1Frames
   outputBaseFileName=sprintf('%3.3d.png',f); % Create index filename
   outputFullFileName=fullfile(outputFolder1,outputBaseFileName);
   imwrite(vid1Frames(:,:,:,f),outputFullFileName,'png');
end

%%
% In this part, the refractive index map is extracted from the video
highIndex = 1.4;        % Manually set based on observation of the video
lowIndex = 1.33;        % Manually set based on observation of hte video
files=dir(outputFolder1);
nx=101; ny=301; nz=201;
indexVolume=1.33*ones(nx,ny,nz);

imageRow1 = 330;            % Manually set based on observation of video
imageRow2 = 630;
imageCol1 = 850;
imageCol2 = 1050;

colorBarRow1 = 80;          % Manually set based on observation of video
colorBarRow2 = 885;
colorBarCol1 = 1384;
colorBarCol2 = 1392;

for i=1:(length(files(:))-2)
    imag_name=files(i).name;
    if length(imag_name)>2
    origRGBImag=imread([outputFolder1 '/' imag_name]);
    grayImage = min(origRGBImag,[],3);
    rgbImage = origRGBImag(imageRow1 : imageRow2, imageCol1 : imageCol2, :);
    colorBarImage = origRGBImag(colorBarRow1 : colorBarRow2, colorBarCol1 : colorBarCol2, :);
    b = colorBarImage(:,:,3);
    storedColorMap = colorBarImage(:,1,:);
    storedColorMap = double(squeeze(storedColorMap)) / 256;
    storedColorMap = flipud(storedColorMap);
    indexedImage = rgb2ind(rgbImage, storedColorMap);
    indexImage = lowIndex + (highIndex - lowIndex) * mat2gray(indexedImage);
    if mod(i,1)==0
        indexVolume(i,:,:)=indexImage;
    end
    end
end
indexVolume=permute(indexVolume,[1,3,2]);
nx=size(indexVolume,1); ny=size(indexVolume,2); nz=size(indexVolume,3);
indexVolume=flip(indexVolume,3);
\end{verbatim}
}
\normalsize

The array of the refractive index must also be scaled to represent the discrete length unit in the out-of-plane dimension, which is different than the discrete length unit used in the images. For this purpose the depth is re-scaled. Borders left over from the images are also cut-out. Finally, the array is padded in each dimension to ensure it exists in a cubic domain. The padded array is saved to file as the highest resolution refractive index.
\tiny
\begin{verbatim}
    %% Re-normalize the index array to desired representation
% Also remove the edge parts extracted from the original images
nxq=ceil(nx*sfx/sfzy);
indexVolume2=1.33*ones(nxq,ny,nz);
dxq=(101/136);  %Ratio of in-plane to out-of-plane slicing.
xq=dxq:dxq:101;

for j=1:ny
    for k=1:nz
        for i=1:nxq
            n=ceil(xq(i));
            indexVolume2(i,j,k)=indexVolume(n,j,k);
        end
    end
end

% Need to get rid of image boundaries
for j=1:ny
    for k=1:nz
        for i=1:nxq
            if (i<23) || (i>123)
                indexVolume2(i,j,k)=1.33;
            end
        end
    end
end

%% Arbitrary translation of array to center
% For the purposes of light scattering, it is helpful to have the
% geometry at the same location with respect to the simulated laser
% and minimum bounding regions. This section cuts of the boundary sections
% and places the yeast cell near the center of the spatial domain. It  also
% adjusts the spatial domain to have the same number of spatial units in
% each dimension by padding the array with background refractive index.



N=300;      % Select 300 since it can include all dimensions
n1=1.33*ones(N,N,N);
for i=1:N
    for j=1:N
        for k=1:N
            if (i>82 && i<219) && (j>50 && j<251) % center the array
                n1(i,j,k)=indexVolume2(i-82,j-50,k);
            end
        end
    end
end

%% Save the output of the yeast cell geometry
save([outputFolder4 'geometry_yeast_n1_' int2str(size(n1,1)) 'x' int2str(size(n1,2)) ...
     'x' int2str(size(n1,3)) '.mat'],'n1');


\end{verbatim}
\normalsize
One additional step to resize the refractive index array is helpful for the cubic lattice in DDSCAT 7.3.3., but is not essential for ANSYS\textsuperscript{\textregistered} Lumerical FDTD. Nonetheless, it is performed for both software to ensure similarity of the structures. In each case, the size refractive index array is reduced to 100 spatial units in each dimension. 
\tiny
\begin{verbatim}
 load geometry_yeast_n1_300x300x300.mat n1
n2=[]; m=0;divd=3;  
for i=1:size(n1,1)
    for j=1:size(n1,2)
        for k=1:size(n1,3)
            if (mod(i,divd)==1)&& (mod(j,divd)==1) && (mod(k,divd)==1)
                m=m+1;
                n2(m)=n1(i,j,k);
            end
        end
    end
end
m=0; n2mod=1.33*ones(floor(size(n1,1)/divd),floor(size(n1,2)/divd),floor(size(n1,3)/divd));
for i=1:floor(size(n1,1)/divd)
    for j=1:floor(size(n1,2)/divd)
        for k=1:floor(size(n1,3)/divd)
            m=m+1;
            n2mod(i,j,k)=n2(m);
        end
    end
end

n1=n2mod;
save(['geometry_yeast_n1_' int2str(size(n1,1)) 'x' int2str(size(n1,2)) 'x' int2str(size(n1,3)) '.mat'],'n1');
 
\end{verbatim}
\normalsize

\section{Refractive index reconstruction}
Refractive index requires unique reconstruction in ANSYS\textsuperscript{\textregistered} Lumerical finite-difference time-domain (FDTD) method, DDSCAT 7.3.3 discrete dipole approximation (DDA), and PyScatman. Each are described in this section. In each simulation domain, the cell size and shape are exactly similar, although each of its dimensions are geometrically scaled by a factor of 0.625$\times$ the dimensions of the cell measured by Habaza \emph{et al.}\cite{habaza2015tomographic} to satisfy memory limits of the WorkStation used for their simulation.

\subsection{DDSCAT 7.3.3 yeast cell reconstruction}
In the DDA, a straightforward file format is used to construct the geometry. Shape data files readable by FORTRAN can be constructed. The shape data file consists of $N$ rows representing independent spatial grid positions that are uniformly distributed, and 5 columns representing the grid index number, the $x$, $y$, and $z$ coordinates, and the specified refractive index material selection as an integer value $M=\{1,...,8\}$. Refractive index nodes can be interpolated to the same positions. Cubic trilinear interpolation is used prior to rounding to nearest refractive index values. Below is an example MATLAB program that exports the reconstructed refractive index array as a shape file.

\tiny{
\begin{verbatim}
% This script takes a local refractive index array, and converts it to a
% format that can be processed by DDSCAT - discrete dipole approximation
% software.
clearvars;
close all;
tic

N=100;
load(['geometry_yeast_n1_' int2str(N) 'x' int2str(N) 'x' int2str(N) '.mat']);
nx=size(n1,1);
ny=size(n1,2);
nz=size(n1,3);

nrk=round(n1,3,'significant');
A1=[1,0,0];A2=[0,1,0];LS=[1,1,1];LO=[-N,-N,-N]/2;

line1 = ' >TARREC   rectangular prism; AX,AY,AZ= %d.4 %d.4 %d.4\n';
line2 = '     %d = NAT\n';
line3 = '  %1.6f  %1.6f  %1.6f = A_1 vector\n';
line4 = '  %1.6f  %1.6f  %1.6f = A_2 vector\n';
line5 = '  %1.6f  %1.6f  %1.6f = lattice spacings (d_x,d_y,d_z)/d\n';
line6 = ' %2.5f %2.5f %2.5f = lattice offset x0(1-3) = (x_TF,y_TF,z_TF)/d for dipole 0 0 0\n';
line7 = '     JA IX IY IZ ICOMP(x,y,z)\n';
lines = '%7.d %3.d %3.d %3.d %2.d %2.d %2.d\n';

fileID = fopen('shape.dat','w');
fprintf(fileID,line1,nx,ny,nz);
fprintf(fileID,line2,numel(nrk));
fprintf(fileID,line3,A1(1),A1(2),A1(3));
fprintf(fileID,line4,A2(1),A2(2),A2(3));
fprintf(fileID,line5,LS(1),LS(2),LS(3));
fprintf(fileID,line6,LO(1),LO(2),LO(3));
fprintf(fileID,line7);
nn=0;
mx_nr=max(nrk(:));
mn_nr=min(nrk(:));
num_nrs=int64(N*(mx_nr-mn_nr));
nrs_cmpnt=[1:(num_nrs+1)];
nrkk =int64((nrk-mn_nr)*N+1);
for k=1:nz
    for n=1:ny
        for m=1:nx
            nn=nn+1;
            fprintf(fileID,lines,nn,m,n,k,nrkk(m,n,k),nrkk(m,n,k),nrkk(m,n,k));
        end
    end
end
fclose(fileID);
toc
\end{verbatim}
}\normalsize
\newpage
\subsection{ANSYS\textsuperscript{\textregistered} Lumerical FDTD yeast cell reconstruction}
ANSYS Lumerical FDTD only accepts geometries specified in GDSII or STL file format. At the writing of this article, the import of complicated geometries defined as an array of refractive index voxels or nodes is not supported. Such points are also a problem for FDTD, since they represent material discontinuities which can cause numerical instability. There are many techniques for constructing a geometry as an STL mesh. Below is an example provided in the MATLAB script used for this manuscript.
\tiny{
\begin{verbatim}
 % This script takes a local refractive index array, and converts it to a
% format that can be processed by ANSYS Lumerical FDTD.
clearvars;
close all;
tic
d=0.105562;             % DDA interdipole spacing returned by DDSCAT 7.3.3.
load('geometry_yeast_n1_100x100x100.mat');
nx=size(n1,1);
ny=size(n1,2);
nz=size(n1,3);
filename='lum_index';
x=(1:nx)*d;      % Measured scaling factor for similar size of DDA
y=(1:ny)*d;      % Measured scaling factor for similar size of DDA
z=(1:nz)*d;      % Measured scaling factor for similar size of DDA
xmn = (5.330884+5.225322)/2.0; ymn=xmn; zmn=xmn; % for centering on the grid
x=x-xmn;
y=y-ymn;
z=z-zmn;
[X,Y,Z]=ndgrid(x,y,z);
for i=1:8
figure; hold on; p=isosurface(X,Y,Z,n1,1.33+(i-1)*0.01); axis equal;
if ~isempty(p.faces) && ~isempty(p.vertices)
    TR = triangulation(p.faces,p.vertices);
    stlwrite(TR,[filename '_n_1_3' int2str(i+2) '.stl']);
end
end
close all;
\end{verbatim}
}
\normalsize
Afterwards, the STL files need to be imported to ANSYS\textsuperscript{\textregistered} Lumerical FDTD with the appropriate scaling factor of microns (0.000001). The exact position is pre-defined by the STL mesh from the provided MATLAB script. Here is a program script to illustrate how each of the STL mesh files can be imported to ANSYS\textsuperscript{\textregistered} Lumerical FDTD. Only run this script once. 

\tiny
\begin{verbatim}
clear;
scalingFactor=0.000001;
k=0;
files = splitstring(dir,endl);    # directory contents in a cell(string) array
for(i=1:length(files)){
 if (findstring(files{i},".stl") != -1)  {  # look for 'fsp' files
    k=k+1;  
  if (fileexists(files{i})) {     # check if the file exists
   ?files{i};               # output file name
    filename=files{i};            # load file
    if (fileexists(filename)){
        stlimport(filename,scalingFactor);
        basename=substring(filename,11,6);
        select("MATLAB STL 2018");
        set("name",basename);          
        mymaterial = addmaterial("(n,k) Material");
        setmaterial(mymaterial,"name",basename);
        setmaterial(basename, "Refractive Index", 1.33+0.01*(k-1)); # enable diagonal anisotropy
        setmaterial(basename, "Imaginary Refractive Index", 0.01);
        select(basename);
        set("material",basename);
    }else
    {
        print('file does not exist, free space region only');
    }  
  }
 }
}    
\end{verbatim}
\normalsize

\newpage
\subsection{PyScatman Yeast cell reconstruction}
PyScatman simply reads in the positions and refractive index values as a four-dimensional array. Refractive index values were similarly cubic interpolated prior to rounding to nearest discrete refractive index values. PyScatman also uses dimensions in microns. 

First the data file is exported to text in MATLAB.
\tiny
\begin{verbatim}
 %   nxs=[50,75,100,150,300]; % optional different mesh sizes
 nxs=100;
for i=1:length(nxs)
   nx=nxs(i); ny=nxs(i); nz=nxs(i);
   filename=sprintf('geometry_yeast_n1_%ix%ix%i.mat',nx,ny,nz);
   load(filename,'er');
    fid=fopen(sprintf('geo_yeast_n1_%ix%ix%i.txt',nx,ny,nz),'w');
   for ii=1:nx
       for jj=1:ny
           for kk=1:nz
                    fprintf(fid,'%6.12f\t',er(ii,jj,kk));
           end
       end
   end
end
close(fid);
\end{verbatim}
\normalsize
Second, the data file is read by Pyton and imported as a \texttt{NumPy} array. The refractive index values are normalized to the background refractive index, such that the scaled refractive index is $n_c/n_b$, where $n_c$ represents the refractive index of the cell, and $n_b$ the refractive index of the background medium. Note that the extinction coefficient (imaginary component) is scaled separately.
\tiny
\begin{verbatim}
sx=100;
sy=100;
sz=100;

file = open("geo_yeast_n1_100.txt", "r")
content = file.readline()
fields = content.split('\t')
file.close()

n1=np.empty(shape=(sx,sy,sz));
n=-1;
for k in range(0,sz):
    for i in range(0,sx):
        for j in range(0,sy):
            n=n+1;
            n1[i,j,k]=fields[n]
# indx_medm=n1-0.33;
indx_medm =1.00+0.75*np.round(np.cdouble(n1-1.33),decimals=2);
indx_medm[indx_medm==1.00]=indx_medm[indx_medm==1.00];
indx_medm[indx_medm!=1.00]=indx_medm[indx_medm!=1.00]+1j*0.01;

\end{verbatim}
\normalsize

\newpage
\section{Comparison of errors for increasing sphere size and different meshing algorithms}

The error of the far-field projection in the Yee-lattice FDTD method slightly decreases as the grid size is increased, suggesting that curvature of the sphere boundary contributes to the numerical error. Although similar to the numerical dispersion imposed by larger unit spacing, this error is more random, and varies significantly for geometries which are smaller with respect to the optical wavelength. In the far-field projection, the error is a slight difference in the angular scattering pattern. These larger systems challenge FDTD and DDA, given the memory required to store the field solutions. Loss (extinction coefficient $\kappa$) was introduced into the final simulation domain, not only for the cell, but for the background medium. Loss decreases the error of the YL-FDTD method far-field projection when compared with Mie theory, especially as the sphere size increases.

\begin{figure}[H]
    \centering
    \includegraphics[width=0.45\linewidth]{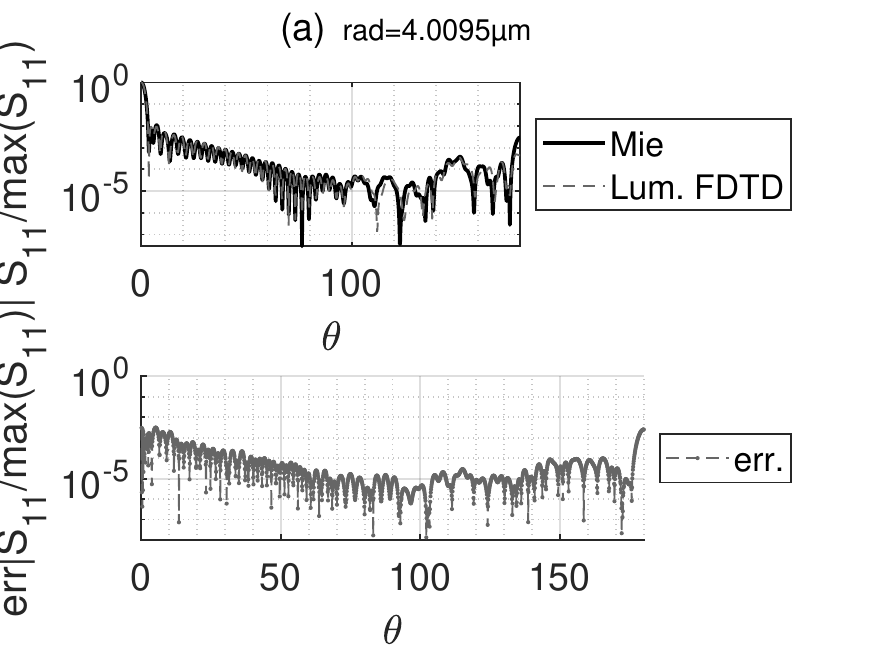}
    \includegraphics[width=0.45\linewidth]{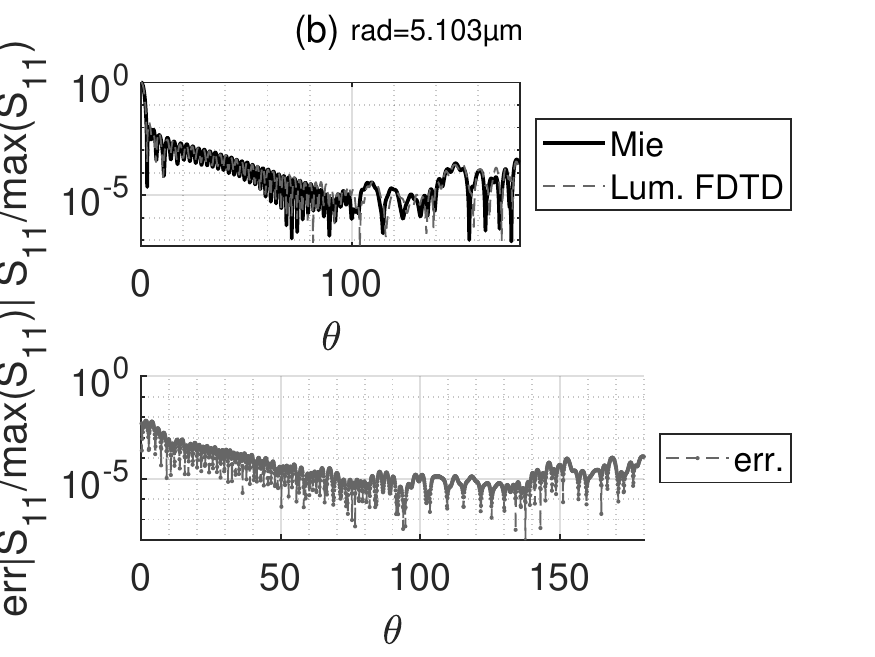}
    \includegraphics[width=0.3\linewidth]{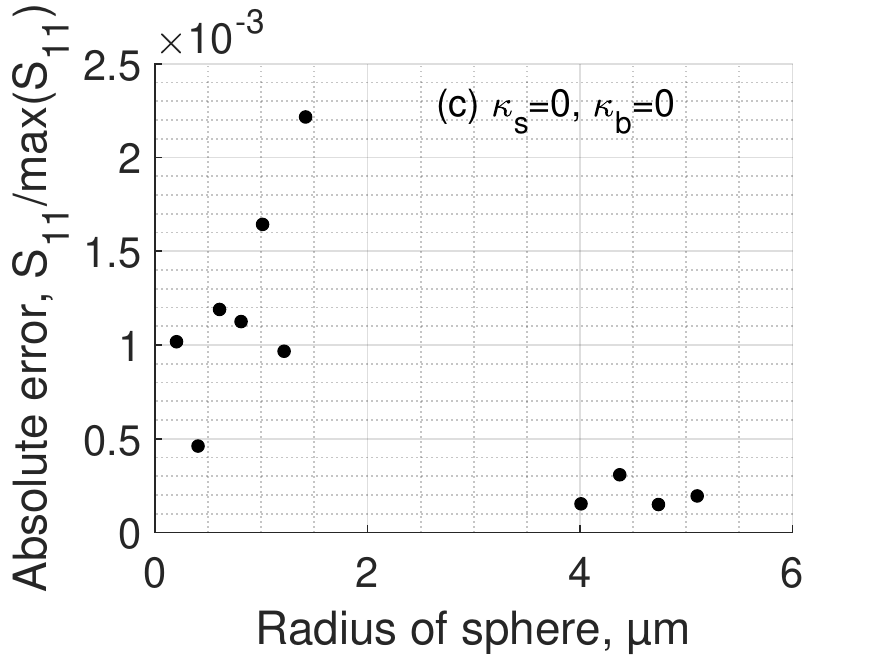}
    \includegraphics[width=0.3\linewidth]{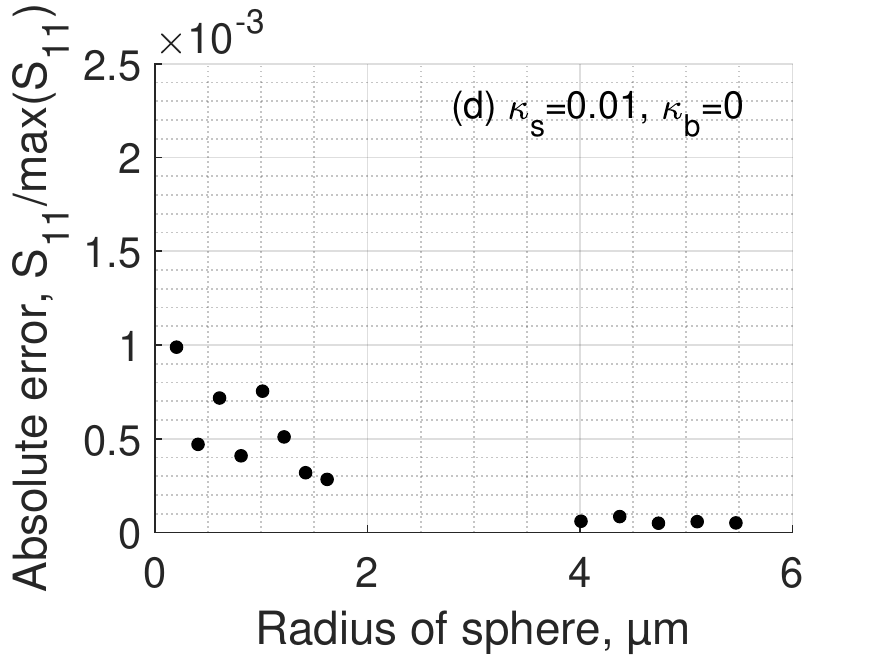}
      \includegraphics[width=0.3\linewidth]{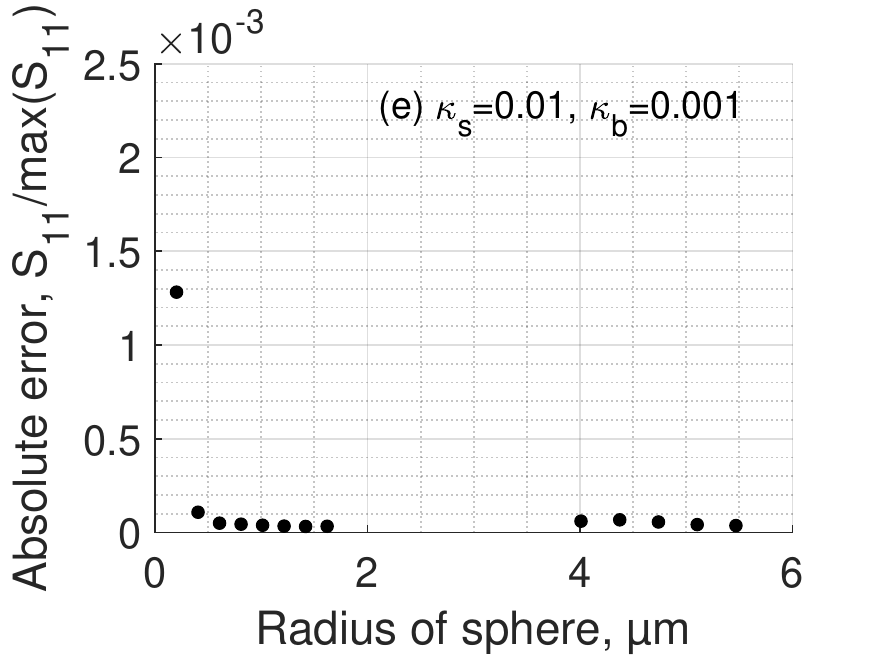}
    \caption{The far-field scattering pattern error (a) with respect to scattering angle $\theta$ for a sphere of radius $r=4.0095$ \si{\micro m} and (b) sphere of radius $r=5.103$ \si{\micro m}. Mean far-field scattering pattern error for (c) lossless spheres in lossless dielectric, where $\kappa_b=\kappa_s=0$; (d) lossy sphere embedded in lossless dielectric $\kappa_b=0$,$\kappa_s=0.01$; and (e) lossy sphere embedded in lossy dielectric $\kappa_b=0.001$,$\kappa_s=0.01$, where the background and sphere extinction coefficients are $\kappa_b$ and $\kappa_s$. The real refractive index of the sphere is $n_s=1.39$, whereas the real refractive index of the background is $n_b=1.33$. }
    \label{fig:enter-label}
\end{figure}

Although a uniform grid was used, the YL-FDTD method error presented in the manuscript was found independent of any type of conformal mapping or non-uniform grids imposed on the sphere structure. This suggests the error is inherent to the Yee algorithm.

\bibliographystyle{unsrt}  
\bibliography{library.bib}

\end{document}